\documentclass[prd,twocolumn,showpacs,superscriptaddress,nofootinbib]{revtex4-2}

\usepackage[T1]{fontenc}
\usepackage[utf8]{inputenc}
\usepackage{natbib}
\usepackage{acronym}
\usepackage{booktabs}
\usepackage{dcolumn}
\usepackage{makecell}
\usepackage{graphicx}
\usepackage{tabularx}
\usepackage{amsmath,amsfonts,amssymb}
\usepackage{mathrsfs}
\usepackage[normalem]{ulem}
\usepackage{multirow}
\usepackage{color}
\usepackage{xcolor}
\usepackage{subcaption}
\usepackage{rotating} 
\usepackage{pifont}
\usepackage{xspace}
\usepackage{aas_macros}

\newcommand{\samplesize}{44~}
\newcommand{\longlived}{20~}
\newcommand{\shortlived}{9~}
\newcommand{\prompt}{9~}
\newcommand{\viscosity}{12~}
\newcommand{\mdisc}{M_{\rm disc}}
\newcommand{\jdisc}{J_{\rm disc}}
\newcommand{\jspec}{j_{\rm spec}}
\newcommand{\cmark}{\ding{51}}%
\newcommand{\xmark}{\ding{55}}%

\newcommand{\is}{{\scriptscriptstyle \mathcal{S}}}

\newcommand{\ie}[0]{i.e.\@\xspace}
\newcommand{\eg}[0]{e.g.\@\xspace}

\newcommand{\Most}{Most et al.\@\xspace}

\definecolor{cyan}{rgb}{0,0.9,0.9}
\definecolor{orange}{rgb}{0.9,0.5,0}
\definecolor{magenta}{rgb}{1,0,1}
\definecolor{purple}{rgb}{0.8,0.4,0.8}
\definecolor{darkgreen}{rgb}{0.0,0.5,0.0}
\definecolor{gray}{rgb}{0.8242,0.8242,0.8242}
\definecolor{cadmiumgreen}{rgb}{0.0, 0.42, 0.24}
\definecolor{olive}{rgb}{0.5, 0.5, 0.0}

\newcommand{\reftab}[1]{{Table~\ref{#1}}}
\newcommand{\refsec}[1]{{Sec.~\ref{#1}}}
\newcommand{\reffig}[1]{{Fig.~\ref{#1}}}
\newcommand{\refapp}[1]{{Appendix~\ref{#1}}}
\newcommand{\refeq}[1]{{Eq.~(\ref{#1})}}
\newcommand{\refpar}[1]{{Par.~\ref{#1}}}

\newcommand{\msun}{\mathrm{~M_\odot}}
\newcommand{\dens}{\mathrm{~g~cm^{-3}}}

\newcommand{\am}{\mathrm{~M_\odot~cm^{2}~s^{-1}}}
\newcommand{\sam}{\mathrm{~cm^{2}~s^{-1}}}
\newcommand{\entr}{k_\mathrm{B}~\mathrm{baryon^{-1}}}
\newcommand{\ye}{Y_{\textnormal{e}}}

\newacro{BNS}{binary neutron star}
\newacro{NS}{neutron star}
\newacro{BH}{black hole}
\newacro{EOS}{equation of state}
\newacroplural{EOS}{equations of state}
\newacro{SD}{standard deviation}
\newacro{AMR}{adaptive mesh refinement}
\newacro{TOV}{Tolman–Oppenheimer–Volkoff}
\newacro{NSE}{nuclear statistical equilibrium}
\newacro{GW}{gravitational wave}
\newacro{GR}{general relativity}
\newacro{NR}{numerical relativity}
\newacro{ISCO}{innermost stable circular orbit}
\newacro{LMSRE}{least mean squared relative error}
\newacro{MHD}{magnetohydrodynamic}
\newacro{GRMHD}{general-relativistic magnetohydrodynamic}
\newacro{MRI}{magnetorotational instability}
\newacro{CCSN}{core-collapse supernova}
\newacroplural{CCSN}[CCSNe]{core-collapse supernovae}
\newacro{GRLES}{general-relativistic large eddy simulations method}

\begin{document}

\title[Accretion disks from binary neutron star mergers]{Geometric and
  thermodynamic characterization of  binary neutron star accretion discs}

\author{Alessandro Camilletti}
\email{Contact e-mail: a.camilletti@unitn.it}
\affiliation{Dipartimento di Fisica, Universit\'{a} di Trento, Via Sommarive 14, 38123 Trento, Italy}
\affiliation{INFN-TIFPA,Trento Institute for Fundamental Physics and Applications, via Sommarive 14, I-38123 Trento, Italy}

\author{Albino Perego}
\affiliation{Dipartimento di Fisica, Universit\'{a} di Trento, Via Sommarive 14, 38123 Trento, Italy}
\affiliation{INFN-TIFPA,Trento Institute for Fundamental Physics and Applications, via Sommarive 14, I-38123 Trento, Italy}

\author{Federico Maria Guercilena}
\affiliation{INFN-TIFPA,Trento Institute for Fundamental Physics and Applications, via Sommarive 14, I-38123 Trento, Italy}
\affiliation{Dipartimento di Fisica, Universit\'{a} di Trento, Via Sommarive 14, 38123 Trento, Italy}

\author{Sebastiano Bernuzzi}
\affiliation{Theoretisch-Physikalisches Institut, Friedrich-Schiller-Universit{\"a}t Jena, 07743, Jena, Germany}

\author{David Radice}
\thanks{Alfred P.~Sloan Fellow}
\affiliation{Institute for Gravitation \& the Cosmos, The Pennsylvania State University, University Park PA 16802, USA}
\affiliation{Department of Physics, The Pennsylvania State University, University Park, PA 16802, USA}
\affiliation{Department of Astronomy \& Astrophysics, The Pennsylvania State University, University Park, PA 16802, USA}

\date{\today}

\begin{abstract}
	Accretion disks formed in binary neutron star mergers play a central role in many astrophysical processes of interest, including the launching of relativistic jets or the ejection of neutron-rich matter hosting heavy element nucleosynthesis. 
	In this work we analyze in detail the properties of accretion disks from \samplesize \emph{ab initio} binary neutron star merger simulations for a large set of nuclear equations of state, binary mass ratios and remnant fates, with the aim of furnishing reliable initial conditions for disk simulations and a comprehensive characterization of their properties. 
	We find that the disks have a significant thermal support, with an aspect ratio decreasing with the mass ratio of the binary from $\sim 0.7$ to 0.3. Even if the disk sample spans a broad range in mass and angular momentum, their ratio is independent from the equation of state and from the mass ratio. This can be traced back to the rotational profile of the disc, characterized by a constant specific angular momentum (as opposed to a Keplerian one) of $3-5 \times 10^{16}\sam$. The profiles of the entropy per baryon and of the electron fraction depend on the mass ratio of the binary. For more symmetric binaries, they follow a sigmoidal distribution as a function of the rest mass density, for which we provide a detailed description and a fit.
	The disk properties discussed in this work can be used as a robust set of initial conditions for future long-term simulations of accretion disks from binary neutron star mergers, posing the basis for a progress in the quantitative study of the outflow properties.
\end{abstract}

\maketitle

\section{Introduction}

Tight systems consisting of two orbiting compact objects eventually merge after a prolonged inspiral phase, during which they lose energy and angular momentum via gravitational radiation \cite{Peters:1963ux,Peters:1964zz}.
In particular, the merger of two \acp{NS}, called a \ac{BNS} merger, results in the formation of a central compact object surrounded by an accretion disk, whose properties depend in a non-trivial way on the binary parameters and on the \ac{EOS} of \ac{NS} matter, see \citep{Shibata:2019wef,Radice:2020ddv,Bernuzzi:2020tgt} for a few recent reviews.
At the end of the inspiral phase, tidal interactions cause the orbiting \acp{NS} to deform, forming spiral arms at the edges of the merging system. In the case of a significantly unequal mass binary, the lighter \ac{NS} is tidally 
destroyed by the more massive one, and a significant fraction of its mass is spread around the more massive one, see e.g. \citep{Rezzolla:2010fd,Hotokezaka:2013mm,Bernuzzi:2014owa,Hotokezaka:2015xka,Bernuzzi:2020txg}.
During the subsequent merger, shocked matter is ejected from the collision interface of the two \acp{NS}. If the total mass of the system is large enough,
a prompt-collapse to a \ac{BH} occurs \citep{Hotokezaka:2011dh,Bauswein:2013jpa,Koppel:2019pys,Bauswein:2020xlt,Kashyap:2021wzs,Perego:2021mkd,Kolsch:2021lub,Tootle:2021umi}, halting matter ejection.
Otherwise, core bounces of the newly-formed massive \ac{NS} remnant expel hot matter in the first few milliseconds that follow the merger, see e.g. \citep{Radice:2018pdn,Perego:2019adq}.
The accretion disk is formed by the gravitationally bound matter expelled during this intricate dynamic.
The later disk evolution is governed by different physical processes, shaping its properties and determining its behavior.
The absorption and emission of neutrinos influence the thermodynamic properties and composition of the disk \citep{Ruffert:1996by,Rosswog:2003rv,Chen:2006rra,Perego:2014fma,Siegel:2017jug,Fujibayashi:2017xsz,Nedora:2020pak}.
Spiral waves \citep{Nedora:2019jhl} and strong magnetic fields \citep{Balbus:1991,Ciolfi:2019fie} can efficiently transport angular momentum during the very first hundreds of milliseconds, 
while on longer, secular timescale the evolution is driven by viscous effects of turbulent magnetic origin \cite{Zurek:1986,Metzger:2008av,Fernandez:2013tya,Just:2014fka,Fujibayashi:2020qda}.
Additionally, the nature of the remnant heavily influences the disk properties. For example, spiral-waves or efficient neutrino irradiation are expected to occur as long as a massive \ac{NS} remnant is present, while the formation of a \ac{BH} remnant causes the innermost and denser part of the disk to be swallowed inside the \ac{BH} horizon, leading to the formation of a lighter torus, see e.g. \citep{Bernuzzi:2020txg,Nedora:2020pak}.

Accretion disks formed in \ac{BNS} mergers are the engine responsible for many relevant processes related to compact binary mergers and to multimessenger astrophysics. It is commonly retained that, in the \ac{BH}-engine scenario, gamma-ray bursts are triggered by the rapid accretion of a magnetized disk into the \ac{BH}, see e.g. \citep{Blandford:1977ds,Blandford:1982di,Lee:1999se,Beloborodov:2008nx,Berger:2014}. Moreover, a relevant portion of the accretion disk, up to  $30-50\%$ of the initial torus mass \citep{Fahlman:2022jkh}, is instead ejected by multiple mechanisms: redistribution of the angular momentum, thermal effects \citep{Metzger:2009xk}, neutrino-driven winds \citep{Perego:2014fma}, magnetic stresses \citep{DeVilliers:2004zz}. This ejected matter is responsible for the nucleosynthesis of heavy elements by means of the so-called rapid neutron capture process (see \citep{Cowan:2019pkx,Perego:2021dpw} and references therein). The radioactive decay of the freshly synthesized, unstable isotopes powers the kilonova transient \citep{Li:1998bw}.
Therefore, the dynamic and thermodynamic properties of the matter inside the disk, together with the mechanisms accountable for the matter accretion and ejection, influence the final abundances of the expelled elements as well as the production of the electromagnetic counterparts associated to \ac{BNS} mergers.

Due to their complexity and high computational costs, only a few previous works have so far simulated \ac{BNS} mergers long enough to account for the evolution of the accretion disks on timescales comparable with the viscous timescale in a fully consistent way \citep{Fujibayashi:2020dvr,Shibata:2021bbj,Radice:2023zlw,Kiuchi:2022nin}.
In many more cases, numerical simulations focusing on the evolution of the accretion disks around a \ac{BH} or a massive \ac{NS} were used to investigate the effects of different mechanisms and the resulting properties of the ejected matter in such a complex scenario \citep{Hawley:2000,Fernandez:2013tya,Just:2014fka,Fernandez:2014bra,Metzger:2014ila,Siegel:2017jug,Fernandez:2018kax,Miller:2019dpt,De:2020jdt,Just:2021cls,Fahlman:2022jkh,Sprouse:2023cdm}.
In these cases, the disks were initialized according to analytical prescriptions that were meant to provide a meaningful description of the disks produced in \ac{BNS} mergers, but that did not directly emerged from merger simulations. 
However, the dynamic and thermodynamic properties of the matter inside these disk lack of an unique analytical description.
As a consequence, the initial conditions in numerical simulations of accretion disks have some degree of arbitrariness. For example, in several cases the disks were initialized using a constant entropy and electron fraction profiles, whose specific values were considered as free parameters, together with the total mass of the disk. 

Despite their relevance, a systematic and comprehensive  characterization of the properties of accretion disks emerging from \ac{BNS} mergers is still missing.
While the properties of the accretion disks resulting from \ac{BH}-\ac{NS} mergers were investigated in \cite{Most:2021ytn}, analysis of the properties of disks emerging from \ac{BNS} merger simulations were so far carried out for limited sets of merger simulations.
In this work, we analyze in detail the geometrical, dynamical and thermodynamic properties of accretion disks from \samplesize \ac{BNS} merger simulations, with the double objective of furnishing a comprehensive characterization of their properties and reliable initial conditions for disk simulations.
In the case of a massive \ac{NS} remnant, the latter and the disk form a continuous structure. However, we separate them by defining a threshold density. In the case of a \ac{BNS} merger collapsing to a \ac{BH}, we consider the disk as the gravitationally bound matter outside the \ac{BH} apparent horizon. It is important to stress that the disk is an evolving system, so its properties depends also on the time at which they are analyzed. In this work, we focus on timescales larger than the formation timescale (a few milliseconds post-merger), but shorter than the secular evolution timescale ($\sim 100$ms).
We observe that some of the prescriptions commonly used to initialize disk simulations do not provide an accurate description of the disk properties as emerging from \ac{BNS} mergers.
In particular, we find that the disks are usually thick, with an aspect ratio decreasing with the mass ratio of the binary,
and with the exception of disks from prompt-collapsed \ac{BNS} mergers, which have a smaller aspect ratio. 
Despite the disk sample spans a broad range in mass and angular momentum, their ratio is independent on the \ac{EOS} and on the mass ratio of the binary. This can be traced back to the rotational profile of the discs, characterized by a constant specific angular momentum. 
We provide fits for the radial and vertical distribution of the rest mass density and of the entropy per baryon and electron fraction distributions with the density.
Our observations might not only be interesting in their own right, but provide a useful and practical way to prescribe initial data for accretion disk simulations with a higher degree of realism.

The paper is structured as follows. In \refsec{sec:code} we summarize the numerical setup and the approximations used to evolve the binaries. The simulation sample is described in \refsec{sec:sample} and the analysis procedure used to define the disk and its properties is illustrated in \refsec{sec:post-processing}. 
The geometrical structure of the discs, i.e. the radial and vertical extensions, the aspect ratio and the half opening angle, are discussed in \refsec{sec:geo_prop}. The mass and angular momentum of the discs, as well as their specific angular momentum and accretion/ejection rates, are the arguments of \refsec{sec:dynamical_properties}. Finally, we investigates the thermodynamic properties, i.e. the electron fraction and the entropy per baryon in \refsec{sec:thermodynamic_prop}. We compare our results with previous numerical simulations of accretion disk in \refsec{sec:discussion}. In this section we also test the rotational model presented in \cite{Galeazzi:2011nn}. The last \refsec{sec:conclusions} conclude the work summarizing the main results.

\section{Methods}\label{sec:methods_and_models}

\subsection{Numerical setup}
\label{sec:code}

All the \ac{BNS} simulations used in this work share the same numerical setup and
microphysics treatment, making their outcome comparable. In particular, we use a
subset of the simulations described in \citet{Perego:2019adq,Endrizzi:2019trv,Nedora:2019jhl,Bernuzzi:2020txg,Nedora:2020pak,Cusinato:2021zin,Perego:2021mkd,Camilletti:2022jms}, part of the CoRe collaboration database \cite{Dietrich:2018phi,Gonzalez:2022mgo}. We briefly describe the employed codes, while the
interested reader can find a detailed description in the aforementioned works.

The numerical methods employed in the works above were implemented in the general
framework provided by the \texttt{EinsteinToolkit} \citep{Loffler:2011ay,
EinsteinToolkit:2022_05}.
They featured a finite-difference scheme to discretize the Einstein's equations,
while
the general relativistic hydrodynamics was handled via the finite-volume
high-resolution shock-capturing code \texttt{WhiskyTHC}
\citep{Radice:2012cu,Radice:2013hxh, Radice:2018xqa}.
All simulations employed the same \texttt{Leakage} + \texttt{M0} scheme
to evolve the changes in composition and energy due to the neutrino interactions
\citep{Galeazzi:2013mia,Radice:2018pdn}. Finally, the time evolution was
performed via a third-order Runge-Kutta scheme, with a constant Courant factor,
which was set based on the speed of light.

The computational domain was covered by seven box-in-box Cartesian grids, where
the resolution of every finer grid was double of the coarser one
\citep{Schnetter:2003rb,Reisswig:2012nc}. The largest refinement level covered a
cube of $\approx 3024~\rm km$ side, while the \acp{NS} and the central object
after the merger were contained in the smaller, most refined level. The
simulations were performed at two or three different grid resolutions. We
characterize each simulation depending on the spacing of the most refined level:
low-resolution (LR), standard-resolution (SR) and high-resolution (HR), with
spacing $\approx 246$, $185$ and $123$ m, respectively. The described domain is
symmetric with respect to the $z = 0$ plane.

The initial conditions of every binary system have been constructed using the
pseudo-spectral elliptic solver \texttt{Lorene} \citep{Gourgoulhon:2000nn},
starting from non-spinning \acp{NS} on quasicircular orbit, with a separation of
$45-50 \; {\rm km}$. The two \acp{NS} were in neutrino-less beta-equilibrium
at a temperature of 0.01 MeV.
 
The \ac{NS} matter was described as a fluid made by neutrons, protons, nuclei,
electrons, positrons, and photons, assuming nuclear statistical equilibrium. The
five finite-temperature, composition-dependent \acp{EOS} which were employed are
broadly compatible with current astrophysical
\citep{Cromartie:2019kug,Miller:2019cac,Riley:2019yda} and nuclear
\citep{Capano:2019eae, Jiang:2019rcw} constraints.
Detailed description of the \acp{EOS} used in this work can be found in
\citet{Logoteta:2020yxf} for the BLh \ac{EOS}, in \citet{Hempel:2009mc} for the
HS(DD2) \ac{EOS}, in \citet{Steiner:2012rk} for the HS(SFHo) \ac{EOS}, in
\citet{daSilvaSchneider:2017jpg} for the SLy4 \ac{EOS} and in
\citet{Lattimer:1991nc} for the LS220 \ac{EOS}. In the following, we will refer
to the second and third ones simply as DD2 and SFHo \acp{EOS}.
Some \acp{EOS} could be disfavored by observational data or theoretical arguments, see \eg
\citet{Tews:2018chv} for LS220 or \citet{Abbott:2018wiz} for DD2. However, the use of several \acp{EOS} allows us to better span present uncertainties.


A total of \viscosity simulations employed the \ac{GRLES} for turbulent viscosity to mimic the effects of large-scale magnetic fields \cite{Radice:2017zta}. Since we observed no significant differences between disks from simulations with or without \ac{GRLES}, we refrain from discussing these twelve cases separately.

\subsection{Simulation sample}\label{sec:sample}

We classify our simulations in three categories: long-lived, where the remnant
does not collapse up to the end of the simulation; short-lived, where the
remnant collapses within the end of the simulation; prompt-collapse, where the
remnant collapse to a \ac{BH} immediately after merger. We identify a prompt
collapse when the minimum of the lapse function decreases monotonically after
merger without any core bounce.

The time indicated as the end of simulation, $t_{\rm end}$, corresponds to the last iteration at which we can retrieve all the data needed for this study (see \refsec{sec:post-processing}). Note that we always express the time with respect to merger. Among the simulations presented in the previous works, we select the ones that last at least 10 ms post-merger for \ac{BNS} merger with long-lived and short-lived
remnant and at least 5 ms post-merger for simulations resulting in a prompt-collapse of the remnant.
With respect to the time of merger, the shorter long-lived simulation in the
sample lasts 10 ms, while the longer lasts 103 ms. Short-lived simulations last
between 16 ms and 36 ms. In the prompt-collapse category, the simulations are as
short as 5 ms and as long as 25 ms. Note that long-lived simulations are not
necessary the longer in our sample and we cannot exclude that a prolonged
evolution would not end up in a \ac{BH} formation. Nevertheless, since the evolution
of the system in the post-merger changes dramatically when the remnant includes
a \ac{NS}, this classification enable us to stress some important differences as
well as genuine similarities between the categories.

The final sample consists of \longlived long-lived, \shortlived short-lived and
\prompt prompt-collapsing \ac{BNS} mergers for a total of \samplesize
simulations, varying in numerical resolution, \ac{EOS}, chirp mass $M_{\rm
chirp}$, mass ratio $q$ and total gravitational mass $M_{\rm tot}$. The mass
ratio of the binaries in the sample spans the range $q \in [1, 1.67]$ and their
total mass is within 2.6 and $3.3~\msun$. Most of the simulations in our sample
are targeted to the \ac{BNS} merger GW170817, with a chirp mass $M_{\rm chirp} = 1.18~\msun$
\citep{TheLIGOScientific:2017qsa}. A set of 6 simulations are targeted to the
\ac{BNS} merger GW190425 with $M_{\rm chirp} = 1.44~\msun$
\citep{Abbott:2020uma}.

\begin{sidewaystable*}
	\centering
	\caption{Simulation sample and the main properties of the disk computed at the
end of the simulation. From left to the right: category according to the fate of
remnant (see \refsec{sec:post-processing}), \ac{EOS}, total initial gravitational mass of the isolated neutron stars $M_{\rm tot}$, mass ratio $q$, inclusion of turbulent viscosity, resolution of the finest grid, time of \ac{BH} formation, end
time of the simulation at which the disk properties are extracted, disk mass
$\mdisc^{\rm end}$, angular momentum $\jdisc^{\rm end}$, specific angular
momentum $\jspec^{\rm end}$, mass-fraction averaged entropy $\langle s^{\rm end}
\rangle$ and electron fraction $\langle \ye^{\rm end} \rangle$. The times are
given from the time of merger. The end time $t_{\rm end}$ indicates
the time at which the last disk can be extracted. The rightmost column collects
the references to the original works where the simulations have been presented.}
	\label{tab:sample}
	\begin{tabular}{cccccc|cc|ccc|ccccc|c}
		\toprule
		Category &EOS &$M_{\rm tot}$ &$q$ &GRLES &res
		&$t_{\rm BH}$ &$t_{\rm end}$
		&$\rm H_{max}$ &aspect &opening
		&$\mdisc^{\rm end}$ &$\jdisc^{\rm end}$ &$\langle \jspec^{\rm end} \rangle$ &$\langle s^{\rm end} \rangle$ &$\langle \ye^{\rm end} \rangle$ &ref\\
		& &$\msun$ & & & &ms &ms &km &ratio &angle &$\msun$ &$\msun^2$ &$\times 10^{16} \sam$ &$\entr$ & & \\
		\hline	long lived &BLh &2.728 &1 &\cmark &SR &\xmark &91 &95 &0.64 &$51^{\circ}$ &0.1328 &1.0831 &3.70 &7.09 &0.12 &\citet{Bernuzzi:2020txg} \\
		long lived &BLh &2.728 &1 &\cmark &HR &\xmark &23 &74 &0.65 &$49^{\circ}$ &0.2081 &1.5398 &3.33 &7.26 &0.15 &\citet{Nedora:2020pak} \\
		long lived &BLh &2.728 &1 &\xmark &HR &\xmark &52 &171 &0.70 &$55^{\circ}$ &0.1139 &0.8875 &3.53 &8.11 &0.16 &\citet{Perego:2020evn} \\
		long lived &BLh &2.730 &1 &\xmark &LR &\xmark &21 &82 &0.63 &$51^{\circ}$ &0.1253 &0.9321 &3.37 &8.11 &0.18 &\citet{Nedora:2020pak} \\
		long lived &BLh &2.730 &1 &\xmark &SR &\xmark &103 &113 &0.67 &$53^{\circ}$ &0.0955 &0.7402 &3.51 &6.13 &0.09 &\citet{Nedora:2020pak} \\
		long lived &BLh &2.765 &1.34 &\xmark &LR &\xmark &41 &134 &0.62 &$52^{\circ}$ &0.2268 &1.8612 &3.74 &7.80 &0.16 &\citet{Nedora:2020pak} \\
		long lived &BLh &2.765 &1.34 &\xmark &SR &\xmark &44 &113 &0.57 &$51^{\circ}$ &0.1664 &1.3716 &3.77 &7.39 &0.14 &\citet{Nedora:2020pak} \\
		long lived &BLh &2.765 &1.34 &\xmark &HR &\xmark &12 &56 &0.48 &$42^{\circ}$ &0.2024 &1.5797 &3.57 &7.17 &0.14 &this work \\
		long lived &BLh &2.803 &1.54 &\xmark &LR &\xmark &40 &74 &0.54 &$45^{\circ}$ &0.2594 &2.2244 &3.92 &6.98 &0.13 &\citet{Nedora:2020pak} \\
		long lived &BLh &2.803 &1.54 &\xmark &HR &\xmark &10 &63 &0.43 &$44^{\circ}$ &0.2454 &2.0294 &3.80 &6.29 &0.11 &\citet{Nedora:2020pak} \\
		long lived &BLh &2.837 &1.66 &\cmark &LR &\xmark &64 &91 &0.52 &$45^{\circ}$ &0.2439 &2.2047 &4.18 &6.77 &0.11 &\citet{Bernuzzi:2020txg} \\
		long lived &BLh &2.837 &1.66 &\cmark &SR &\xmark &19 &78 &0.43 &$43^{\circ}$ &0.2650 &2.2929 &3.99 &6.86 &0.12 &\citet{Bernuzzi:2020txg} \\
		long lived &BLh &2.837 &1.66 &\cmark &HR &\xmark &15 &65 &0.40 &$43^{\circ}$ &0.2538 &2.1919 &3.99 &6.25 &0.10 &\citet{Bernuzzi:2020txg} \\
		long lived &DD2 &2.728 &1 &\xmark &LR &\xmark &38 &70 &0.57 &$45^{\circ}$ &0.2117 &1.7140 &3.67 &6.79 &0.13 &\citet{Nedora:2019jhl} \\
		long lived &DD2 &2.728 &1 &\xmark &SR &\xmark &92 &96 &0.60 &$48^{\circ}$ &0.1811 &1.4914 &3.73 &6.00 &0.10 &\citet{Perego:2019adq} \\
		long lived &DD2 &2.732 &1.10 &\xmark &LR &\xmark &41 &49 &0.51 &$45^{\circ}$ &0.2349 &1.9157 &3.68 &6.26 &0.12 &\citet{Cusinato:2021zin} \\
		long lived &DD2 &2.733 &1.11 &\xmark &LR &\xmark &26 &57 &0.55 &$44^{\circ}$ &0.2582 &2.0782 &3.63 &6.72 &0.14 &\citet{Cusinato:2021zin} \\
		long lived &DD2 &2.740 &1.19 &\xmark &LR &\xmark &28 &76 &0.57 &$44^{\circ}$ &0.2557 &2.0764 &3.68 &7.12 &0.14 &\citet{Cusinato:2021zin} \\
		long lived &DD2 &2.742 &1.20 &\xmark &LR &\xmark &36 &69 &0.57 &$44^{\circ}$ &0.2530 &2.0934 &3.74 &6.75 &0.14 &\citet{Nedora:2020pak} \\
		long lived &DD2 &2.880 &1.67 &\cmark &SR &\xmark &30 &117 &0.50 &$48^{\circ}$ &0.2753 &2.5445 &4.26 &7.24 &0.14 &this work \\
		\hline
		short lived &LS220 &2.728 &1 &\cmark &LR &18 &27 &111 &0.58 &$49^{\circ}$ &0.1605 &1.2549 &3.48 &7.74 &0.14 &\citet{Nedora:2019jhl} \\
		short lived &LS220 &2.728 &1 &\cmark &SR &13 &20 &122 &0.64 &$51^{\circ}$ &0.0502 &0.3901 &3.45 &8.60 &0.19 &\citet{Nedora:2019jhl} \\
		short lived &LS220 &2.728 &1 &\xmark &LR &17 &32 &130 &0.68 &$55^{\circ}$ &0.0697 &0.5450 &3.47 &7.68 &0.15 &\citet{Nedora:2019jhl} \\
		short lived &LS220 &2.728 &1 &\xmark &SR &15 &36 &97 &0.65 &$52^{\circ}$ &0.0631 &0.4951 &3.49 &7.21 &0.13 &\citet{Nedora:2019jhl} \\
		short lived &LS220 &2.737 &1.16 &\cmark &SR &20 &26 &151 &0.66 &$55^{\circ}$ &0.1238 &0.9947 &3.57 &8.39 &0.18 &\citet{Nedora:2020pak} \\
		short lived &LS220 &2.781 &1.43 &\cmark &LR &15 &19 &103 &0.58 &$51^{\circ}$ &0.1855 &1.6314 &3.93 &7.77 &0.15 &\citet{Nedora:2020pak} \\
		short lived &SFHo &2.735 &1.13 &\xmark &SR &11 &16 &185 &0.73 &$54^{\circ}$ &0.0775 &0.5902 &3.38 &9.74 &0.23 &\citet{Nedora:2020pak} \\
		short lived &SLy4 &2.728 &1 &\xmark &SR &13 &21 &264 &0.66 &$56^{\circ}$ &0.0417 &0.3136 &3.34 &11.1 &0.24 &\citet{Endrizzi:2019trv} \\
		short lived &SLy4 &2.735 &1.13 &\xmark &SR &12 &19 &229 &0.61 &$55^{\circ}$ &0.0687 &0.5282 &3.42 &9.87 &0.23 &\citet{Nedora:2020pak} \\
		\hline
		prompt &BLh &3.307 &1.12 &\xmark &LR &0 &10 &15 &0.33 &$27^{\circ}$ &0.0010 &0.0081 &3.79 &6.50 &0.06 &\citet{Camilletti:2022jms} \\
		prompt &BLh &3.307 &1.12 &\xmark &SR &0 &12 &11 &0.26 &$23^{\circ}$ &0.0005 &0.0044 &3.74 &6.65 &0.05 &\citet{Camilletti:2022jms} \\
		prompt &BLh &3.322 &1.18 &\xmark &LR &0 &7 &5 &0.17 &$15^{\circ}$ &0.0030 &0.0252 &3.77 &4.84 &0.05 &\citet{Camilletti:2022jms} \\
		prompt &LS220 &2.837 &1.66 &\cmark &LR &1 &14 &58 &0.30 &$28^{\circ}$ &0.1185 &1.2242 &4.63 &4.29 &0.05 &\citet{Bernuzzi:2020txg} \\
		prompt &LS220 &2.837 &1.66 &\xmark &LR &1 &14 &58 &0.30 &$30^{\circ}$ &0.1155 &1.1760 &4.57 &4.33 &0.05 &\citet{Bernuzzi:2020txg} \\
		prompt &SFHo &2.837 &1.66 &\cmark &SR &1 &25 &48 &0.36 &$30^{\circ}$ &0.0906 &0.8925 &4.40 &5.72 &0.07 &\citet{Bernuzzi:2020txg} \\
		prompt &SFHo &3.322 &1.18 &\xmark &SR &0 &7 &13 &0.27 &$20^{\circ}$ &0.0005 &0.0043 &3.87 &7.63 &0.06 &\citet{Camilletti:2022jms} \\
		prompt &SFHo &3.351 &1.33 &\xmark &SR &0 &5 &5 &0.19 &$14^{\circ}$ &0.0109 &0.0957 &3.89 &3.61 &0.04 &\citet{Camilletti:2022jms} \\
		prompt &SLy4 &3.322 &1.18 &\xmark &SR &0 &6 &6 &0.18 &$18^{\circ}$ &0.0005 &0.0041 &3.80 &7.06 &0.09 &\citet{Camilletti:2022jms} \\
		\bottomrule
	\end{tabular}				
\vspace{-.4\textheight}
\end{sidewaystable*}  
\reftab{tab:sample} lists the main properties of the simulated binaries categorized according to the fate of remnant.

\subsection{Data analysis procedure}\label{sec:post-processing}
To exploit the intrinsic symmetries of the system, we adopt cylindrical
coordinates with the axial direction aligned to the rotational axis of the
binary. The radial and axial extensions of the cylinder are $\approx 1181~\rm km$, characterized by a constant spacing of $\approx 148$ m up to a distance of $295$ km and a logarithmic spacing for the successive 200 grid points along both coordinates. The azimuthal angle $\phi$ is divided in 62 sections of $\approx 5.71^\circ$. The cylindrical coordinates $r$, $z$ and $\phi$ refer to the cell centers of the resulting grid. The hydrodynamic variables are linearly interpolated from the seven Cartesian refinement levels into the cylindrical grid, using values in the most refined level available around each cell center of the cylindrical grid.

\begin{figure*}
	\includegraphics{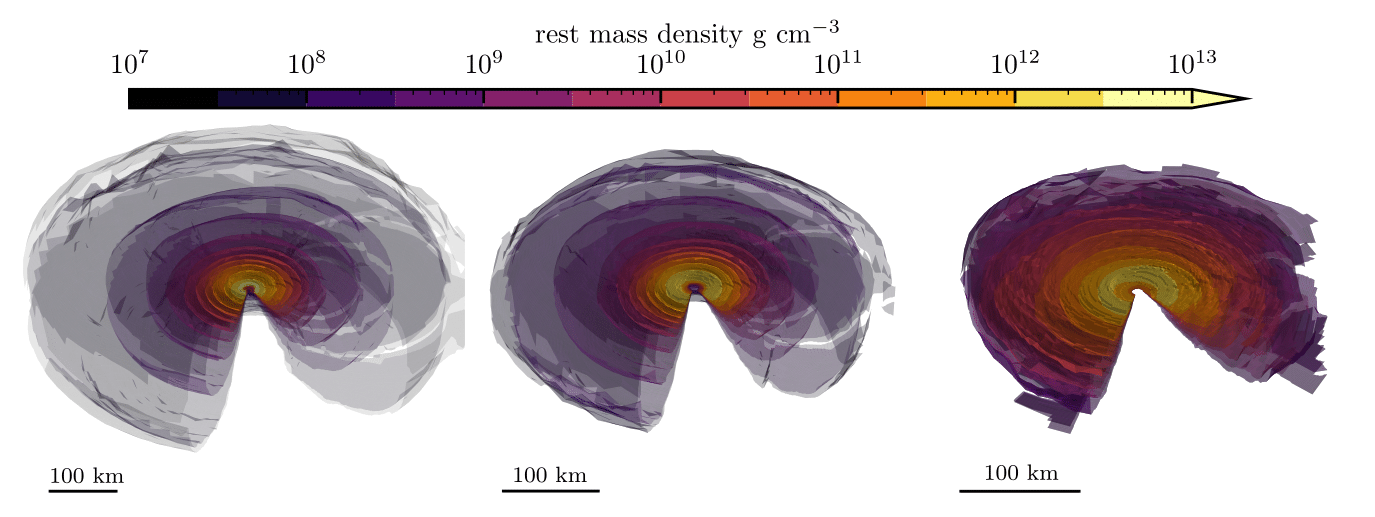}
	\caption{Isodensity surfaces for three representative simulations taken at $t_{\rm end}$. Left: disk from a long-lived \ac{BNS} merger, as obtained from the equal mass, HR simulation with the BLh \ac{EOS} (without turbulent viscosity). Center: disk from a short-lived \ac{BNS} merger, as obtained from the equal mass, SR simulation with the LS220 \ac{EOS} (without turbulent viscosity). Right: disk from a prompt-collapsed \ac{BNS} merger, as obtained from the SR simulation with SFHo \ac{EOS} (with turbulent viscosity).}
	\label{fig:3Dshape}
\end{figure*}

We define the disk as the bound matter which rest mass density is $\rho \leq
10^{13}~\dens$ \citep[see \eg][]{Shibata:2017xdx} and $\rho \geq \rho_{\rm
min}$, where $\rho_{\rm min}$ is such that the disk mass is 95\% of the total baryon mass enclosed in the grid with $\rho \le 10^{13}~\dens$.
This method allows us to circumvent the need for defining a cut-off that depends on a minimum rest mass density. Such a definition could overly rely on this arbitrary choice, potentially influencing the total mass of the disk. At the same time, it prevents the inclusion of regions with extremely low densities in the disk, which might be susceptible to numerical artifacts.
If the \ac{BNS} collapse to a \ac{BH}, we remove matter in the space-time region whose lapse function is less than 0.3. The ejecta is removed from the disk according to the geodesic criterion, \ie fluid elements with $|u_t| \geq c$, where $u_t$ is the time-component of the four-velocity. The regions that satisfy the previous requirements but are disconnected from the main disk body are finally removed. Despite these regions are usually small, they can artificially affect the geometrical properties of the disk and must be removed.
The isodensity surfaces of the disk obtained from this procedure are showed in \reffig{fig:3Dshape} for three representative simulations.

The disk mass is computed as the general-relativistic volume integral of the conserved baryon mass density inside the volume of the torus:
\begin{equation}\label{eq:disc_mass}
	\mdisc = \int_{\rm disc} \sqrt \gamma \rho W ~rdr d\phi dz ~,
\end{equation}
where $\rho$ is the baryon rest mass density, $W$ is the Lorentz factor of the fluid and $\sqrt{\gamma}$ is the determinant of the 3-metric. Analogously, assuming symmetry with respect to the rotational axis, we define the disc angular momentum as the general-relativistic volume integral of the baryon angular momentum density along the azimuthal direction $j = \rho h W^2
\tilde v_{\phi}$, \ie
\begin{equation}\label{eq:disc_am}
	\jdisc = \int_{\rm disc} \sqrt \gamma \rho h W^2 \tilde v_{\phi} ~rdr d\phi dz ~,
\end{equation}
where $h$ is the fluid specific enthalpy and $\tilde v_{\phi}$ is the advective angular velocity in the azimuthal direction. We recall that, in cylindrical coordinates, the advective azimuthal velocity is related to the Cartesian components of the fluid Eulerian velocity $v_i$ as $\tilde v_{\phi} = \alpha \left( x v_y - y v_x \right) - \beta_{\phi}$, where $\alpha$ is the lapse function and $\beta_{\phi}$ is the low index $\phi$-component of the shift vector. Note that the assumption of an axial symmetric space-time is approximately satisfied after a relaxation phase. The specific angular momentum, \ie the angular momentum per unit of mass, is the ratio between the baryon angular momentum density and rest mass density $j / \rho$.

We define the aspect ratio at every azimuthal angle $\phi_0$ of the cylindrical grid as the ratio between the maximal radial and vertical extensions of the disk in the $\phi = \phi_0$ plane.
An average over $\phi$ is then performed to obtain the aspect ratio of the disc. Analogously, on each $\phi = \phi_0$ plane, the half opening angle is defined as the arcotangent of the ratio between the maximum vertical extension and the radial distance at which this maximum is found. An average over $\phi$ is then performed to obtain the half opening angle of the disc.

The flux of baryon mass is computed as $\mathbf{f} = \rho W \mathbf{\tilde v}$ (note that here and in the following we define the advective velocity $\mathbf{\tilde v} = \alpha \mathbf{v} - \boldsymbol{\beta}$). We derive the accretion / ejection rate $\dot M$ across a spherical surface as the flux integral:
\begin{equation}\label{eq:accretion}
	\dot M = 2\int_{0}^{\pi/2} \int_0^{2\phi} \sqrt{\gamma_S(\theta,\phi)} f^r(\theta,\phi) ~d\theta d\phi
\end{equation}
where $\theta$, $\phi$ are the polar and azimuthal coordinates on the spherical surface, $\gamma_S$ is the pull-back of the spatial metric on it, and $f^r$ is the radial component of the baryon mass flux.
Note that, when computing the flux, we interpolate the latter quantities on a spherical grid.

In many occasions, we perform a non-linear least-square fit between two
hydrodynamic variables $x$ and $y$. For example, in
\refsec{sec:thermodynamic_prop} we fit the distribution of the entropy and of
the electron fraction with respect to the rest mass density. If $\{\mu\}$ is a set of parameters of the fitting relation $y(x,\{\mu\})$, the determination of $\{\mu\}$ is performed by minimizing the residuals weighted by the
mass fraction, \ie $m_f|y - y(x,\{\mu\})|$, where $m_f = dm / \sum dm$ is the mass
fraction and $dm = \rho r \Delta r \Delta z \Delta \phi$ is the baryon rest mass
in each grid cell. 
In most cases we found similarities between the fit performed on simulations belonging to the same category, \ie prompt, short or long-lived. 
We characterize each category $\small{\mathcal{C}}$ with a representative set of parameters $\{\langle \mu_\mathcal{C} \rangle\}$ by computing a weighted average of the parameters $\mu_{\is}$ obtained from the fit on each simulation $\mathcal{S}$ in the category $\mathcal{C}$:
\begin{equation}\label{eq:average}
	\langle \mu_\mathcal{C} \rangle = \frac{\sum\limits_{\is \in \mathcal{C}} \mu_{\is} w_{\is}}{\sum\limits_{\is \in \mathcal{C}} w_{\is}}~,
\end{equation}
where  $w_\mathcal{\scriptscriptstyle S} = 1/\sigma_\mathcal{\scriptscriptstyle S}^2$, and $1/\sigma_\mathcal{\scriptscriptstyle S}^2$ are the 1-$\sigma$ standard deviations of the fitted parameter (as estimated by the least-squares method).
The error $\sigma_{\langle \mu_\mathcal{C} \rangle}$ on each averaged parameter $\langle \mu_\mathcal{C} \rangle$ is computed as
\begin{equation}\label{eq:error}
	\sigma_{\langle \mu_\mathcal{C} \rangle} = \sqrt{ \frac{\sum\limits_{\is \in \mathcal{C}} w_{\is}(\langle \mu_\mathcal{C} \rangle - \mu_{\is})^2}{\sum\limits_{\is \in \mathcal{C}} w_{\is}} }~.
\end{equation}

If not stated differently, we discuss the disk properties at $t_{\rm end}$ defined in \refsec{sec:sample}. Indeed, we are mostly interested in describing the disks properties once a steady configuration has been reached.

\section{Results}

\subsection{Geometric properties}
\label{sec:geo_prop}

\subsubsection*{Spatial extension.}
At the reference time $t_{\rm end}$, the radial extension of the discs in our sample spans
the range $47 - 736$ km and the vertical extension can be as small as 7 km and
as large as 390 km, where smaller values are found for lighter discs. In
particular, the simulations targeted to the \ac{BNS} merger GW190425 undergo
prompt-collapse due to the high total mass of the system, resulting in lighter
and smaller discs. These disks are characterized by radii between $47-85$ km and
vertical extension between $7-22$ km.

\begin{figure*}
	\includegraphics{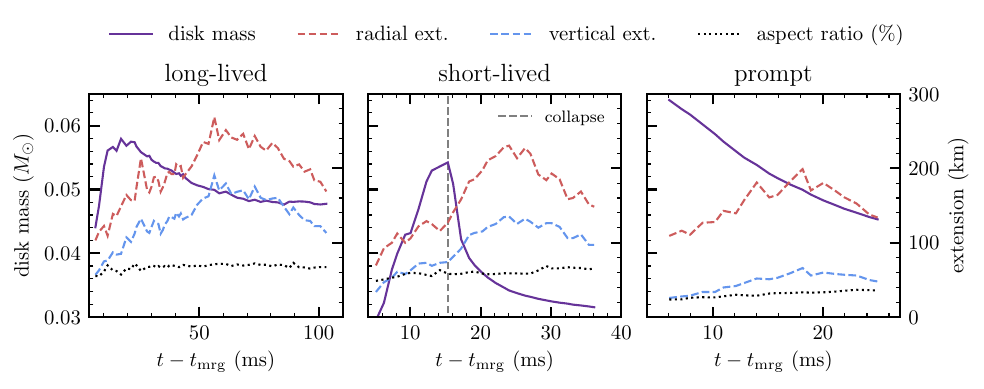}
	\caption{Disk mass (left axis), radial extension, vertical extension and aspect ratio in percentage (right axis) for a sample of SR simulations in each category. Left panel: long-lived \ac{BNS} merger obtained from the $q=1$, simulation with the BLh \ac{EOS} (without turbulent viscosity). Central panel: short-lived \ac{BNS} merger obtained from the $q=1$ simulation with the LS220 \ac{EOS} (without turbulent viscosity). Right panel: prompt-collapsed \ac{BNS} merger obtained from the $q=1.66$ simulation with the SFHo \ac{EOS} and turbulent viscosity.}
	\label{fig:mass_extent}
\end{figure*}
Regarding the time evolution of the spatial extents (see \reffig{fig:mass_extent}), we note that after an initial expansion reflecting the disk formation, the accretion onto the central object and the ejection of matter contribute to reduce the disk volume, decreasing both the radial and vertical extensions. However, this change in volume essentially does not affect the shape of the disk and in particular its aspect ratio (see below).


\subsubsection*{Aspect ratio and opening angle.}
\label{sec:aspect_ratio}
\begin{figure}
	\includegraphics{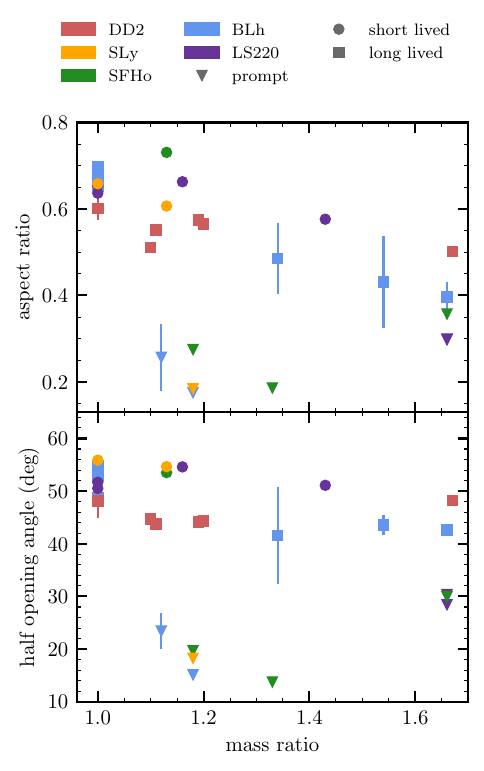}
	\caption{Relation between the aspect ratio (top panel) and the half opening angle (bottom panel) with the mass ratio of the binary. Colors represent the \ac{EOS} while markers label the fate of remnant. Values are taken at the last timestep of the highest-resolution simulation available for each \ac{BNS} model. Errors are estimated as the difference between the two higher resolutions available.}
	\label{fig:aspect_ratio}
\end{figure}
The top panel of \reffig{fig:aspect_ratio} shows the relation between the aspect ratio of the disk and the mass ratio of the binary.
All the discs in our simulation sample are considerably thick, with an aspect ratio between 0.2 and 0.7. This clearly indicate that, in addition to the rotational support, remnant disks are characterized by a significant thermal support. According to scaling relations related to the vertical structure of the disk, the aspect ratio can be estimated by the ratio between the sound and the rotational speed inside the disk. For a few representative simulations, we have verified that the ratio between these two speeds is $\sim 0.2-0.4$ across the orbital plane and within the innermost 100km, in good qualitative agreement with our aspect ratio results.
The disks from prompt-collapsed \ac{BNS} mergers are located in the lower region of the plot, below an aspect ratio $\sim 0.4$, while disks from long-lived and short-lived simulations span a broader range and are usually thicker. In general, the aspect ratio of the disks from long and short-lived simulations tends to decrease with $q$ from a maximum of $\sim 0.7$ in the equal mass cases to a minimum of $0.4$ for very asymmetric binaries, $q \gtrsim 1.6$. Moreover, for those simulations, the data suggest that the aspect ratio decreases faster for softer \ac{EOS}, but more unequal \ac{BNS} merger simulations are needed to clearly asses this trend.
Both these trends are likely related to the effects of tidal interactions in the disk formation process, since they are more pronounced for stiffer \acp{EOS} and higher mass ratios. Indeed, tidal interactions expel matter from the central object predominantly towards the orbital plane, increasing the disk extend in this direction and therefore reducing the aspect ratio.

\begin{figure}
	\includegraphics{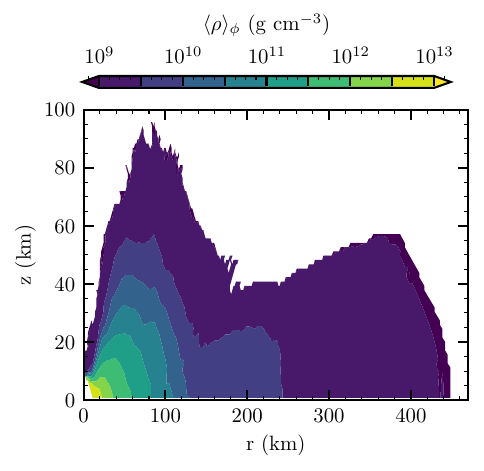}
	\caption{Mass fraction $\phi$-averaged rest mass density distribution on the
    $rz$-plane for a $q=1.66$ long-lived \ac{BNS} merger HR simulation with BLh
    \ac{EOS} (without turbulent viscosity). The distribution is taken at the last available timestep.}
	\label{fig:rmd_rz_high_q}
\end{figure}

The bottom panel of \reffig{fig:aspect_ratio} shows the opening angle as a function of the mass ratio. Again, disks from prompt-collapsed \ac{BNS} mergers are in the low region of the plot, where the half opening angle is $\sim 10-30^\circ$. Instead, the disks from simulations in the long and short-lived category are wider and have a higher half opening angle that goes from $42^\circ$ to $56^\circ$.
Differently from the aspect ratio, the trend of the half opening angle with $q$ is less pronounced and is more ore less constant for $q \gtrapprox 1.3$. Indeed, while the matter at large latitude is expelled by shocks and remnant bounces during the merger, in the case of high mass ratio binaries, the disk tends to include a tail at large radii formed by the tidal disruption of the lighter \ac{NS}, as shown in \reffig{fig:rmd_rz_high_q}. Since the half opening angle is computed from the ratio between the maximum height and the radial distance at which the maximum height is found, it is not affected by the presence of a tail at larger radii, which instead affects the aspect ratio.

We note that our data do not allows us to infer the presence of any trends of aspect ratio or half opening angle vs. mass ratio regarding models that undergo prompt-collapse. Additional simulations covering a larger set of binary configurations would be needed to investigate this point.

\subsection{Dynamical properties}\label{sec:dynamical_properties}

\subsubsection*{Minimum rest mass density.}\label{par:rmd_min}

The post-processing procedure described in \refsec{sec:post-processing} implies
that every disk in our simulation sample has a different minimum rest mass
density $\rho_{\rm min}$, which also changes in time. To give a reference,
general values of $\rho_{\rm min}$ (taken at the last timestep of each
simulation) are in the range $10^7 - 4 \times 10^{10} \dens$ with a geometric mean of $2 \times 10^8 \dens$.

Lower values of $\rho_{\rm min}$ are reached in light disks produced by
prompt-collapsed \ac{BNS} mergers: since their maximum density attains the
typical value of $\sim 10^{12} \dens$, such disks extend towards lower densities
to meet the requirement of containing 95\% of the total baryon mass of the
system.

\subsubsection*{Disc mass and angular momentum.}
\begin{figure}
	\includegraphics{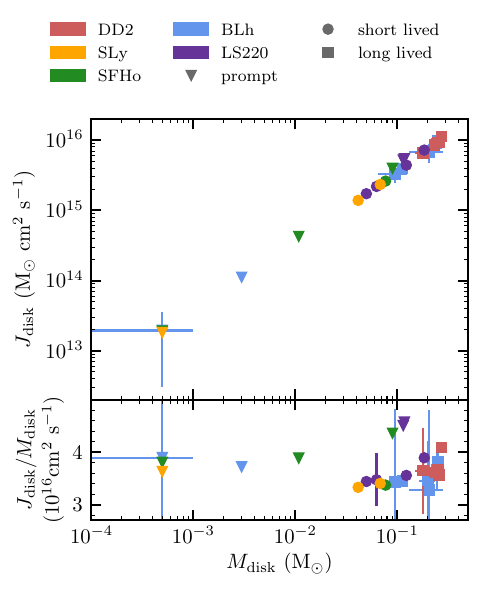}
	\caption{Disc mass $\mdisc$ and angular momentum $\jdisc$ as defined in
\refeq{eq:disc_mass} and \refeq{eq:disc_am} and their ratio for each \ac{BNS}
merger model at the highest resolution available in our sample. Values are taken
at the end of the simulation. Errors are estimated as the difference between the
two higher resolutions available. Colors (markers) represent the \ac{EOS} (fate
of the remnant).}
	\label{fig:mass_angular_momentum}
\end{figure}

The mass and angular momentum of the disks span a broad range of values, going
from $5 \times 10^{-4}$ to $0.3 \msun$ for the mass, and from $2 \times 10^{13}$
to $10^{16} \am$ for the angular momentum.
Disc mass and angular momentum are summarized in the top panel of
\reffig{fig:mass_angular_momentum} and in \reftab{tab:sample}. We underline that
the values of $\mdisc$ and $\jdisc$ that we report are $\approx 5\%$ smaller from
those found in the papers in which the simulations considered here were presented first.
This is due to slight differences in the definition of disk in
those works with respect to the present one (see \refsec{sec:post-processing}).
From the figure it can be seen that $\jdisc$ and $\mdisc$ are distributed along a power law (\ie a
linear relation in log scale). Moreover, their ratio $\jdisc/\mdisc$ is almost
constant over 3 orders of magnitude in $\mdisc$ spanning the tight range $3.3 - 4.6 \times 10^{16} \sam$. This generalizes previous findings only related to disks produced in prompt-collapsing simulations targeted to GW190425 \citep{Camilletti:2022jms}.

\begin{figure}
	\includegraphics{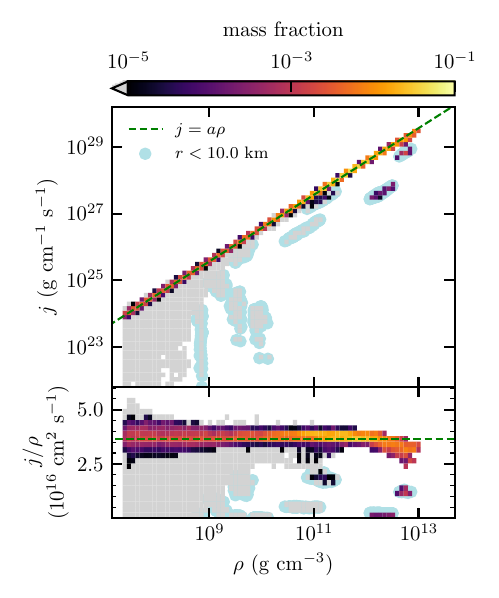}
	\caption{Mass weighted histogram of the angular momentum density and the rest
mass density of the disk (top) and of the specific angular momentum (bottom),
obtained from the long-lived equal mass merger HR simulation with BLh \ac{EOS}
(without turbulent viscosity). The color scale represents the fraction of
$\mdisc$ in every bin. When the mass fraction is smaller than $10^{-5}$ the bin
is gray. Bins related to fluid elements at radii smaller than 10 km are
highlighted in light blue.}
	\label{fig:rmd_amd}
\end{figure}
Similarly to the disk mass and angular momentum, also the respective integrands,
\ie the rest mass density and angular momentum density, exhibit a power law
relation as depicted in \reffig{fig:rmd_amd}. Only the fluid elements near the
remnant (highlighted in light blue) deviate from the trend, but their mass
fraction is $\lessapprox 10^{-2} - 10^{-3}$ smaller than the mass fraction of
the volume elements that follow the power law behavior.
\label{sec:sam}
\begin{table}
	\caption{
	Weighted averages and uncertainties, computing according to \refeq{eq:average} and \refeq{eq:error}, respectively, for the parameters obtained from the least square fits of the $j=j(\rho)$ relation \refeq{eq:linear_sam} (left) and of the $j_G = j_G(\Omega)$ relation \refeq{eq:galeazzi_jw} (right, see \refsec{subsec:sam_discussion}), separately for each simulation sub-sample (long-lived, short-lived, prompt). All the fits have been carried out at $t_{\rm end}$ for each simulation in the sample.}
	\begin{tabular}{c|c|ccc}
		\toprule
		&$\langle a \rangle$ &$\langle \Omega_c \rangle$ &$\langle R_0 \rangle$ &$\langle \alpha \rangle$ \\
		&($10^{16}~\rm cm^2~s^{-1}$) &(kHz) &(km) &\\
		\hline
		long-lived  &$3.9 \pm 0.3$ &$19 \pm 4$ &$13 \pm 2$ &$-1.03 \pm 0.04$ \\
		short-lived &$3.5 \pm 0.3$ &$31 \pm 11$ &$8 \pm 2$ &$-1.02 \pm 0.01$ \\
		prompt      &$4.7 \pm 0.3$ &$45 \pm 13$ &$8 \pm 1$ &$-1.10 \pm 0.01$ \\
		\bottomrule
	\end{tabular}
	\label{tab:rmd_amd_galeazzi_fit}
\end{table}
\reffig{fig:rmd_amd} suggest a power-law relation $j = a \rho^\gamma$ between the rest mass density and the angular momentum density. We have found that the power-law exponents $\gamma$ are approximately 1 for all the simulations on which the fit has been performed.
This implies that the specific angular momentum (\ie the ratio between the angular momentum density $j$ and the rest mass density $\rho$) is almost constant over the disc, as shown in the bottom panel of \reffig{fig:rmd_amd}. 
We characterize the proportionality between $j$ and $\rho$ by fitting a linear relation 
\begin{equation}\label{eq:linear_sam}
	j = a \rho ~,	
\end{equation}
minimizing the relative mean square error.
\begin{figure}
	\includegraphics{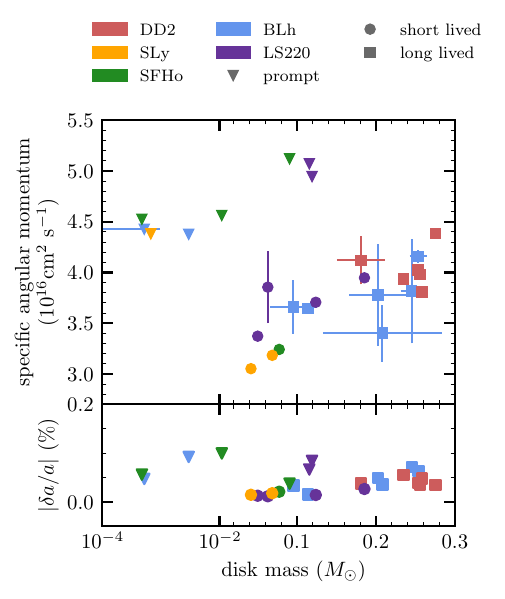}
	\caption{Specific angular momentum obtained from the linear fit of the angular momentum density as a function of the rest mass density (Eq.~\ref{eq:linear_sam}). Values are taken at $t_{\rm end}$ and for the simulation with highest resolution for each \ac{BNS} merger model. Errors are estimated as the difference between the two highest resolutions available. The bottom panel shows the one standard deviation relative error on the fitted slope. Colors (markers) represent the \ac{EOS} (the fate of the remnant).}
	\label{fig:rmd_amd_fit_lin}
\end{figure}
\reffig{fig:rmd_amd_fit_lin} shows the specific angular momentum obtained by the linear fit for all the simulations in our sample, which is contained in the tight range $\sim 3-5 \times 10^{16} \sam$. This is consistent with both the mass-weighted averages $\langle \jspec^{\rm end} \rangle$ in \reftab{tab:sample} and the bottom panel of \reffig{fig:mass_angular_momentum}, despite $\jdisc/\mdisc$ being the ratio of
integrated quantities. Note that a mass-weighted average gives more relevance to the fluid elements with higher mass-fraction, \ie  in the disk regions at higher densities. The fit performed here is not weighted by  the mass fraction and we believe it is a better estimator for the specific angular momentum of the whole disc. 
We find that the specific angular momentum increases with the disk mass and with the mass ratio of the binary. 
Furthermore, disks originating from prompt-collapsed
\ac{BNS} mergers possess specific angular momentum that falls within the higher bounds of the aforementioned range, with values between 4 and $5 \times 10^{16} \sam$.
In these kinds of mergers, the disk is mostly composed of tidally ejected material, which is expelled with larger angular momentum during the late inspiral and from the merging \acp{NS}.
In the long and short lived cases, and especially in the equal mass mergers, the disk is mostly formed by matter expelled after the merger through shocks and bounces originating from the remnant. Several mechanisms, including gravitational wave emission, are very effective in removing or redistributing angular momentum from the remnant. This has a quantitative impact on the specific angular momentum of the matter that forms the disc.

The values of the fitted parameter averaged among the simulations in each category are reported in \reftab{tab:rmd_amd_galeazzi_fit}. Note that \reftab{tab:rmd_amd_galeazzi_fit} also contains the results of the fits for the parameters discussed in \refsec{subsec:sam_discussion}.

\subsubsection*{Accretion rate}
\begin{figure}
	\includegraphics{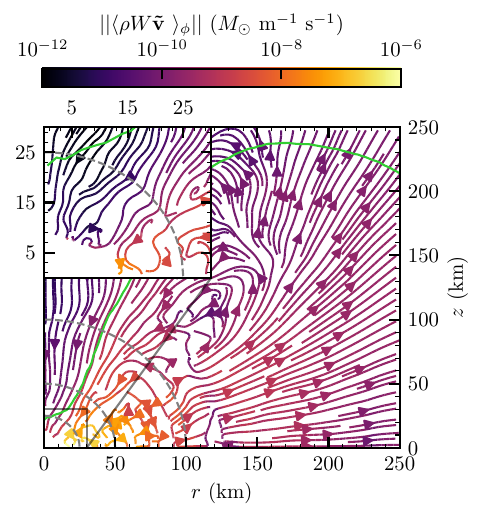}
	\caption{$\phi$-averaged flow lines of matter for the long-lived equal mass merger HR simulation with BLh \ac{EOS} (without turbulent viscosity), taken at $\sim 52 \rm ms$ post-merger. The green line is the disk border. The gray dashed circles represent spherical surfaces of 25, 50 and 100 km radius, on which the azimuthal distributions of $\dot M$ displayed in \reffig{fig:Mdot_angle_chosen} are computed.}
	\label{fig:flux_long_lived}
\end{figure}
\reffig{fig:flux_long_lived} show the $\phi$-averaged flow lines of conserved rest mass density $\langle \rho W \mathbf{\tilde v}
\rangle_{\phi}$ for a long-lived equal mass \ac{BNS} merger at the last available timestep, where $\mathbf{\tilde v}$ is the advective velocity. At latitudes below $45^\circ$ and within $1/3$ of the total radial extension of the disk (\eg 100 km in \reffig{fig:flux_long_lived}),
the flux of matter is disordered, with alternating regions of inflow and outflow. However at larger radii the conserved mass density flux is mostly outgoing. The accretion/ejection rates of this matter flux are computed across spherical surfaces according to \refeq{eq:accretion}, without imposing a limit on the minimum and maximum rest mass density of the fluid elements considered.
\begin{figure*}
	\includegraphics{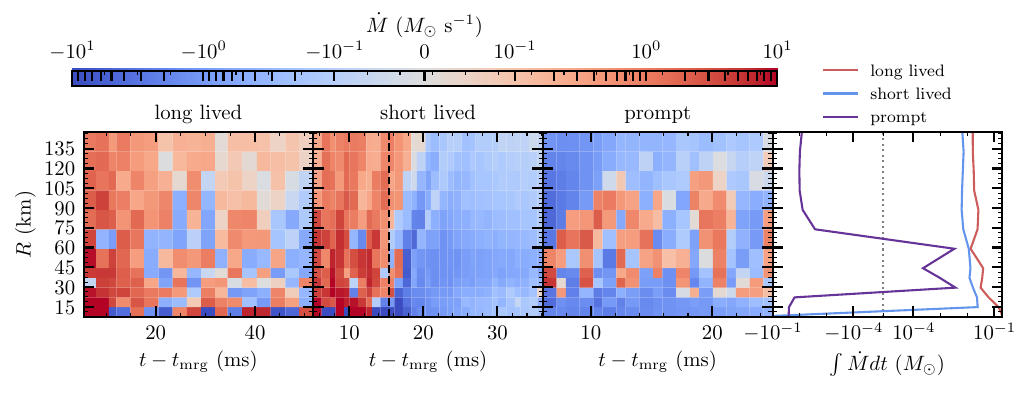}
	\caption{
	Time evolution of the total accretion / ejection rate across spherical surfaces of fixed coordinate radius $R$. From left to right: long-lived, short-lived and prompt-collapsed \ac{BNS} mergers chosen from the simulations sample, \ie  the equal mass merger HR simulation with BLh \ac{EOS} (without turbulent viscosity), the equal mass merger SR simulation with LS220 \ac{EOS} (without turbulent viscosity) and the SR simulation with SFHo \ac{EOS} and $q=1.66$. The last plot on the right shows the total (\ie time integrated) mass crossing each spherical surface for the three scenarios. The vertical dashed line in the short-lived plot indicates the \ac{BH} formation time.}
	\label{fig:Mdot_R_chosen}
\end{figure*}
As summarized in \reffig{fig:Mdot_R_chosen}, the absolute value of the accretion/ejection rates $|\dot M|$ can reach $\sim 10 \msun~\rm s^{-1}$. In the first ten milliseconds after merger, disks with a \ac{NS} in the center have a persisting outflow of matter across every sphere of radius between 10 and 140 km, with peaks above $10~\msun~\rm s^{-1}$. The outflow decreases with time and can alternate with episodes of inflow in the inner region of the disc, where the flux of the conserved mass density is more disordered.
Nevertheless, the total $\dot M$ remain positive at every radius with values $\sim 10 - 100 ~\msun~\rm s^{-1}$ (see rightmost top panel).

Before \ac{BH} formation, the accretion/ejection rate in \ac{BNS} mergers with short-lived remnant have a behavior similar to the long-lived ones, characterized by a net ejection of matter for sufficiently large radii.
This persistent outflow is due to a combination of multiple mechanism. In the very first milliseconds after merger, the outflow is due to the tidal torques in the late inspiral and to the expanding shocks produced at merger and originating from the bouncing remnant in the center.
On longer timescales, the absorption of neutrinos and the spiral waves from the central \ac{NS} further contribute to the outflow.
A significant accretion onto the central object only occurs after the \ac{BH} formation (vertical dashed line). In the
prompt-collapsed \ac{BNS} mergers the total outflow strongly depends on the radius of the spherical surface in consideration. The central \ac{BH} of prompt-collapsed \ac{BNS} mergers always accretes matter at small radii but a net outflow is possible in the inner regions of the disc. At larger radii the trend inverts again with a net inflow of matter.

\begin{figure*}
	\includegraphics{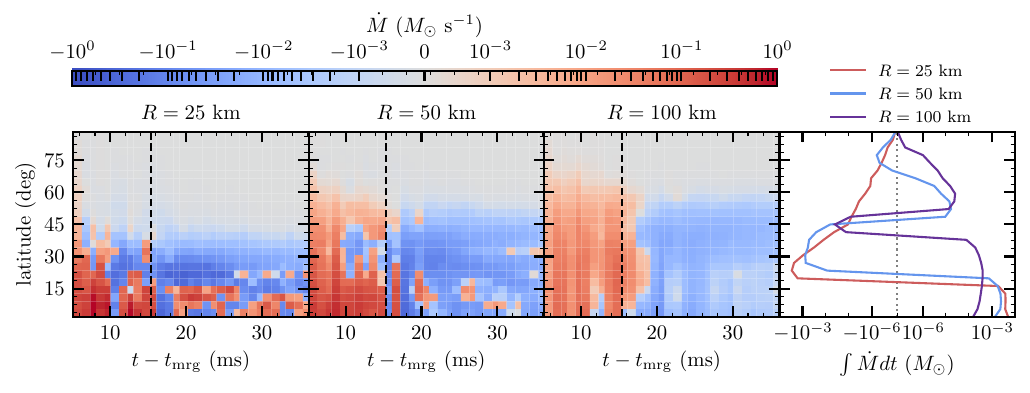}
	\caption{
		Time evolution of the angular distribution of the accretion/ejection rate across spherical surfaces of radii $R = 25,~50,~100$ km (from left to right) for the same short-lived simulation of \reffig{fig:Mdot_R_chosen}. The last plot on the right shows the total mass crossing each spherical surface at the various angles. The vertical dashed lines indicate the \ac{BH} formation time. Note that the color coded scale is different with respect to \reffig{fig:Mdot_R_chosen}.}
	\label{fig:Mdot_angle_chosen}
\end{figure*}
The polar distribution of the accretion/ejection rate across spherical surfaces of radius 25, 50 and 100 km, integrated along the azimuthal coordinate, is shown in \reffig{fig:Mdot_angle_chosen} for the same short-lived simulation of \reffig{fig:Mdot_R_chosen}. At very early times ($t-t_{\rm mrg} \sim 1{\rm ms}$) the shock-heated matter expelled from the central \ac{NS} spreads to all latitudes and, in the successive 10 ms, the matter forming the disk is expelled at latitudes $\lesssim 30^\circ$. Near the central \ac{NS} (left panel), after the first 10 ms accretion dominates at latitudes $\gtrsim 30^\circ$, while episodes of inflow and outflow alternate closer to the orbital plane. Indeed, the ratio between the radial and azimuthal velocity in the equatorial plane is $\sim 10^{-2}$ and the orbital period is $\sim 1.5~\rm ms$, suggesting that the radial flux can change sign rapidly. 
Increasing the distance from the central \ac{NS} (middle and right panels), neutrino absorption and nuclear recombination release energy in the regions at intermediate latitude $\sim 30-60^\circ$, where the density and temperature are lower (see \eg figure~16 of \citet{Perego:2014fma}), increasing the outflow at such latitudes.
At larger distances (center and right panels) and until \ac{BH} formation, a significant ejection of matter characterizes a broad range of latitude, up to  $60^\circ$. Afterwards, but with a delay increasing with the radial distance, the ejection turns into an accretion flow at all latitudes.
At any spherical surface the higher values of outflow rate are reached at lower latitudes where most of the mass is concentrated (see the rightmost panel of \reffig{fig:Mdot_angle_chosen}). 
\ac{BNS} mergers with different fate of the remnant display similar behavior in the polar distribution of $\dot M$, with ejection on broader angles at larger radii and a transition to an inflow after a \ac{BH} is formed.

\subsubsection*{Spatial distribution of the rest mass density.}
\label{par:rmd_rz}
In the following, we discuss the results of an empirical fitting procedure applied to the mass-weighted $\phi$-average of the rest mass density as a function of the radial and height coordinates, $\rho(r,z)$. Note that the coordinates in the simulations are gauge dependent and the results presented here are qualitative in nature. However, the disks extend into a region where the gravitational pull of the central object is rather weak, and the gauge conditions employed in our simulations tend towards geodesic coordinates in these conditions. We can therefore expect to be able to provide a useful
description of the mass distribution despite its gauge dependence. Indeed, we show that our procedure produces satisfactory results even when applied to simulations with different \ac{EOS}, mass ratio and fate.

We observe that the rest mass density distribution in the $rz$ plane can be approximated by the product of three terms: the maximum of the rest mass density at $z=0$, $\max \rho(r,z=0)$, its rescaled radial distribution, $\rho_0(r) \equiv \rho(r,z=0) / \max \rho(r,z=0)$, and its rescaled rest mass density distribution along $z$, \ie  $\rho_{\bar r}(z) \equiv
\rho(\bar r,z)/\max (\rho(\bar r,z))$. In the following, we describe the fitting procedure for $\rho_0(r)$ and $\rho_r(z)$ in detail.

We fit $\rho_0(r)$ with the same relation used in \citet{Camilletti:2022jms} apart from a normalization, \ie a Gaussian centered on a radius $r_0$ and of variance $\sigma_0$ smoothly joined to a power-law decay
\begin{equation}\label{eq:rmd0_rz}
\rho_0(r) = 
\begin{cases}
\exp\left[-(r-r_0)^2/\sigma_0^2 \right] &r \le r_* ~,\\
\exp \left[-(r_*-r_0)^2/\sigma_0^2 \right] (r/r_*)^{-\alpha} &r > r_* ~,
\end{cases}
\end{equation}
where $\alpha = 2r_* (r_*-r_0)/\sigma_0^2$ and the threshold $r_*$ is a free parameter. We note that for long-lived remnants, $r$ is always greater than $r_*$ due to the $10^{13} \dens$ threshold on the rest mass density.
In this case we fit only the power law decay as $\rho_0(r) = \max\{(r/r_*)^{-\alpha}, 1\}$, with $\alpha$ being a free parameter in the fit.

Inspired by the analytic solution for an isothermal not-self-gravitating disk, the rescaled rest mass density distribution along $z$, \ie  $\rho_{\bar r}(z) $ at every fixed $\bar r$ in the grid, is fitted using a Gaussian continuously joined to a decaying exponential
\begin{equation}\label{eq:rmd_rz_fit}
\rho_{\bar r}(z) = 
\begin{cases}
\exp \left[-z^2/H(\bar r)^2 \right] &z \le z_*(\bar r) ~, \\
\exp \left[ -z_*(\bar r)^2/H(\bar r)^2 \right] e^{-\beta} &z > z_*(\bar r) ~,
\end{cases}
\end{equation}
where $\beta = (z-z_*(\bar r))/z_0(\bar r)$. The scale-height of the disk, $H(\bar r)$, and the $z_0(\bar r)$ and $r_*(\bar r)$ parameters are then fitted as functions of radius with the following relations:
\begin{subequations}\label{eq:rmd_rz_fit_param}
	\begin{equation}\label{eq:rmd_rz_fit_sigma}
	H(r) = mr + p ~,
	\end{equation}
	{
		\addtolength{\belowdisplayskip}{-5ex}
		\addtolength{\belowdisplayshortskip}{-5ex}
		\addtolength{\abovedisplayskip}{-5ex}
		\addtolength{\abovedisplayshortskip}{-5ex}
		\begin{equation}\label{eq:rmd_rz_fit_z0}
		z_0(r) = ar^2 + br + c ~,
		\end{equation}
	}
	\begin{equation}\label{eq:rmd_rz_fit_zs}
	z_*(r) = A \log_{10}(r/B) ~.
\end{equation}
\label{eq:sigmazz0_r}
\end{subequations}
Eqs.~(\ref{eq:rmd_rz_fit_sigma}), (\ref{eq:rmd_rz_fit_z0}) and (\ref{eq:rmd_rz_fit_zs}) have no direct physical interpretation and are modeled ad-hoc on the data.
In \reftab{tab:rmd_rz_parameters} of \refapp{app:rmd_rz}, we report the values of the various parameters obtained from the fitting procedure described above separately for each of our simulations. Simulations with short-lived and prompt-collapsed remnant have values of $r_0$, $r_*$ ad $\sigma_0$ in the ranges $12-127~\rm km$, $15-40~\rm km$ and $6-27~\rm km$, respectively. In the case of long-lived simulations, $r_*$ varies from $15$ to $24~\rm km$ and $\alpha$ from $3$ to $5$.
Regarding the parameters in \refeq{eq:rmd_rz_fit_sigma}, $m$ ($p$) varies between 0.07 (-7 km) and 1 (10 km). Note that $H$ given by \refeq{eq:rmd_rz_fit_sigma} is negative for $r < -p/m$ when $p < 0$. This imposes a minimum radius at which the fitting procedure can effectively approximate the rest mass density distribution of the disk. The minimum and maximum of the parameters $a$, $b$ and $c$ of \refeq{eq:rmd_rz_fit_z0} are -0.003 and 0.015, -0.9 and 0.6, -5 and 18, respectively for each parameter. Finally, $A$ and $B$ of \refeq{eq:rmd_rz_fit_zs} varies from 4 to 93 and from 0.01 to 8.5, respectively.
As can be seen from the corner plots in \reffig{fig:rmd_rz_corr} of \refapp{app:rmd_rz}, some of the parameters introduced in Eqs.~(\ref{eq:sigmazz0_r}) could be correlated, and these correlations could possibly be exploited to reduce the number of parameters.

The complete rest mass density as a function or $r$ and $z$ is finally obtained as $\rho(r,z) = \max \rho(r,z=0) \rho_0(r) \rho_r(z)$ inserting Eqs.~(\ref{eq:sigmazz0_r}) into Eq.~\eqref{eq:rmd_rz_fit}. Since this procedure only involves rescaled quantities, the maximum of the rest mass density on the $xy$ plane, $\max \rho(r,z=0)$, can be chosen to obtain the desired disk mass once the other parameters have been fixed.
\begin{figure}
	\centering
	\includegraphics{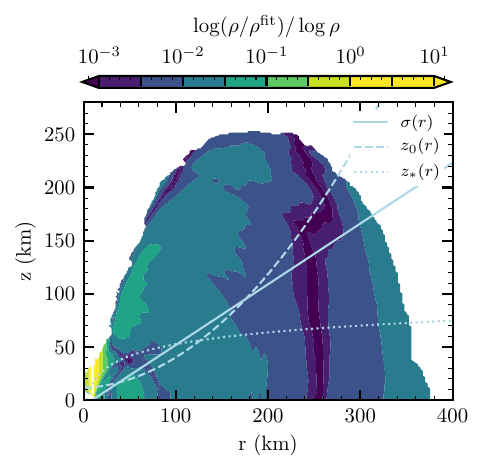}
	\caption{Relative difference between the logarithms of the mass-weighted $\phi$-average rest mass density and the fit discussed in \refsec{par:rmd_rz}, for the equal mass long-lived \ac{BNS} merger HR simulation with BLh \ac{EOS} (without turbulent viscosity), at the end of the simulation. The computation of the relative difference has been limited to the region occupied by the disc. Solid, dashed and dotted lines represent, respectively, the quantities $H(r)$, $z_0(r)$ and $z_*(r)$ obtained by fitting the parameters of \refeq{eq:rmd_rz_fit} using \refeq{eq:rmd_rz_fit_param}.}
	\label{fig:rmd_rz_reldiff}
\end{figure}
\reffig{fig:rmd_rz_reldiff} shows the relative difference between the logarithms of the mass-fraction $\phi$-averaged rest mass density and the results of the fitting procedure. The fit is able to capture the rest mass density distribution with average relative error of $\sim 10^{-2}$, excepting for a narrow region near the remnant where it can reach a factor of the order of 10.
The geometric mean of the relative difference of the logarithms $\log(\rho/\rho^{\rm fit}) / \log(\rho)$, averaged over the simulation sample is contained in the range $0.3 - 6~\times 10^{-2}$. Furthermore the accuracy of the fit is slightly poorer for high-$q$ models than for low-$q$ ones. Overall, this indicates that the fit is able to appropriately describe the rest mass density distribution of most of the disk over a variety of configurations.

\subsection{Thermodynamic properties}
\label{sec:thermodynamic_prop}

\subsubsection*{Entropy}
\label{sec:entropy}

We find that the distribution of the entropy in the disk changes significantly depending on the mass ratio of the binary. In particular, $q \approx 1.3$ seems to be a threshold between two different regimes as also found in \citet{Perego:2021mkd}. This motivate us to separate the discussion in small and high mass ratio cases.
\newline
\paragraph*{Small mass ratio ($q \lessapprox 1.3$)}

\begin{figure}
	\includegraphics{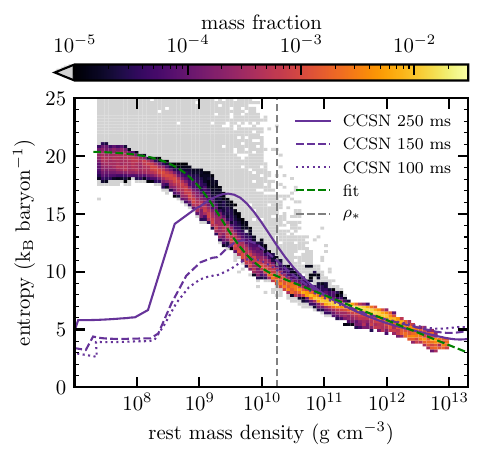}
	\caption{Histogram of the distribution of the baryon mass vs. rest mass density and entropy of the disk for the representative equal mass long-lived \ac{BNS} merger HR simulation with BLh \ac{EOS} (without turbulent viscosity), taken at the last available timestep. The fit with \refeq{eq:rmd_entr_fit} is shown using a green dashed line. The purple lines are the density - entropy distributions from the \ac{CCSN} simulation discussed in \ref{sec:discussion} at different post-bounce time.}
	\label{fig:rmd_entr_chosen}
\end{figure}

\reffig{fig:rmd_entr_chosen} shows the typical distribution of the matter inside the disk in terms of the rest mass density and entropy per baryon. Most of the disc, in terms of mass, has entropy between 4 and 8 $\entr$. These values are found in the high density region with $\rho \sim 10^{10} - 10^{13} \dens$. This region may be only very roughly be regarded as isentropic. At lower density the entropy per baryon increases and reaches a plateau around $15 - 20 \entr$ at $\rho \lesssim 10^9 \dens$. The distribution produced by \ac{BNS} mergers characterized by the prompt-collapse of the remnant shows a similar behavior, but the final plateau in entropy occurs at lower values of $\sim 10 - 15 \entr$.

The entropy distribution in the disk is determined by the dynamics that follows the first milliseconds after merger and the initial disk formation. Matter inside the inspiraling \acp{NS} has very low entropy. The subsequent dynamics produces shocks that increase the entropy in many different ways. First, there is the production of a shock at merger, at the collisional interface between the two merging \acp{NS}. Despite the large speed of the collision, the resulting shock is weak due to the large sound speed of nuclear matter ($c_s \gtrsim 0.2 c$ for matter around saturation density). Under these conditions, the jump in entropy per baryon can be estimated as
\begin{equation}
  \Delta s \sim \frac{\Gamma (\Gamma + 1)}{12} \left( \frac{\Delta v}{c_s}
  \right)^3\,,
\end{equation}
where $\Gamma $ is the adiabatic index ($\sim 5/3$ for non-relativistic nucleons, while $\sim 4/3$ if the equation of state is dominated by relativistic electrons or photons) and $\Delta v$ the variation of the speed at the shock front. Considering that the speed variation cannot exceed the orbital speed at merger ($v_{\rm orb} \sim 0.4c$), $\Delta s \lesssim 3 \entr$. Secondly, the bounces of the central object produce radial sound waves that become shock waves at the edges of the forming remnant, expelling shock-heated dynamical ejecta, with a typical entropy between 10 and 15 $\entr$. At the same time, they also expel shock-heated matter from the collisional interface, which collides with the faster and rotating spiral arms formed by the tidal tails of the two \acp{NS}. The latter are characterized by initially unshocked matter at lower entropy that gets shocked by the collision with the warmer and slower matter in the disc. The typical sound speed inside the disk decreases down to 0.02c for $\rho \sim 10^{8} \dens$, while $\Delta v$ across the shock front is a decreasing fraction of $ v \sim v_{\rm orb} \propto R^{-1}$ such that $\Delta v / c_s$ is of the order of a few and $\Delta s \sim 10 \entr$. The overall effect is a monotonic increase of the specific entropy. After a few orbits, the action of the shocks ceases.
Until gravitational collapse to a \ac{BH}, the aftermath of the merger is marked by the propagation of spiral waves originating from the central \ac{NS}. The propagation of these waves is adiabatic and any change in the entropy distribution occurs solely due to expansion, which brings matter with $s \sim 20-25 \entr$ to densities below $\sim 10^9 \dens$. Note that in this discussion we are not considering fluid elements in the disk with mass fraction below $10^{-5}$, where the entropy can be much higher.

Based on the data found in our sample, we propose a functional relation between the entropy per baryon $s$ and the rest mass density $\rho$, consisting in an arcotangent smoothly joined to a logarithm:
\begin{equation}
  \label{eq:rmd_entr_fit}
  s(\rho)=
  \begin{cases}
    s_0 - \bar s \arctan(\rho / \rho_0)
    & \rho \leq \rho_* \,,
      \vspace{12pt}\\
    \begin{split}
      s_0 - \bar s &\arctan(\rho_* / \rho_0) \\
                   &- \frac{\ln 10}{\rho_0/\rho_* + \rho_*/\rho_0} s_0 \log(\rho/\rho_*)  
    \end{split}
    & \rho > \rho_* \,.
  \end{cases}
\end{equation}
The parameters $s_0$, $\bar s$, $\rho_0$ and $\rho_*$ are obtained by a non-linear least-squares fit with residuals weighted by the mass fraction $m_f$ of the fluid elements.
\begin{table}
	\caption{Same as in \reftab{tab:rmd_amd_galeazzi_fit}, but for the parameters obtained from the least square fits of the $s=s(\rho)$ relation \refeq{eq:rmd_entr_fit}. The column "No. of sims" indicates the number of simulations in each category over which the average is performed.}
	\label{tab:rmd_entr_fit_par}
	\begin{tabular}{cc|cccc}
		\toprule
		Category & No. &$s_0$ &$\bar s$ &$\rho_0$ &$\rho_*$\\
		& of sims &\multicolumn{2}{c}{$(\rm k_B~baryon^{-1})$} &\multicolumn{2}{c}{($\times 10^{10} \rm g~cm^{-3}$)}\\
		\hline
		long-lived  &14 &$17 \pm 2$ &$6 \pm 1$ &$0.3 \pm 0.1$ &$3 \pm 1$\\
		short-lived &8  &$17 \pm 1$ &$6 \pm 1$ &$1.1 \pm 0.3$  &$11 \pm 3$\\
		prompt      &4  &$10 \pm 5$ &$6 \pm 2$ &$0.2 \pm 0.1$ &$4.2 \pm 0.1$\\
		\bottomrule
	\end{tabular}
\end{table}
In \reftab{tab:rmd_entr_fit_par} we report the averaged values of the parameters appearing in Eq.~\ref{eq:rmd_entr_fit} for each simulation category. The parameters $s_0$ and $\bar s$, \ie the entropy of the plateau at low density and the magnitude of the jump in the transition region, are comparable among simulations in the long-lived and short-lived categories. The central density $\rho_0$, around which the transition from low entropy to the entropy plateau occurs, differs by almost one order of magnitude instead.
\begin{figure}
	\centering
	\includegraphics{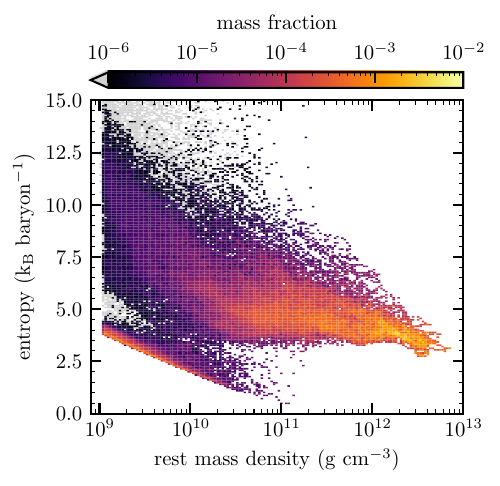}
	\caption{Rest mass density and entropy per baryon histogram for the prompt-collapsed \ac{BNS} merger SR simulation with $q = 1.66$ and SFHo \ac{EOS}, at 11 ms after merger. The fluid elements in the low density ($\sim 10^{9} - 10^{11}~\dens$) and low entropy ($\lessapprox 5~\entr$) belong to the tidal component of the disc.}
	\label{fig:rmd_entr_highq}
\end{figure}
\newline
\paragraph*{High mass ratio ($q \gtrapprox 1.3$).}

As the mass ratio increases, the lower-mass star in the system is more and more likely to be tidally disrupted at the time of merger. This disrupted matter is then flung outwards, mostly along the orbital plane. By this process, the proportion of the tidally-ejected mass forming the disk increases with respect to the portion ejected by shock heating. At mass ratios $q \gtrsim 1.3$, part of the tidally-ejected matter in the early post-merger forms a component separated from the bulk of the disc. This component is clearly visible in the low entropy, low density region of \reffig{fig:rmd_entr_highq}. This component has entropy per baryon below $5 \entr$ and density of $\rho \lesssim 10^{11} \dens$. Furthermore we have observed that, as the system evolves, $\lessapprox 10\%$ of this tidal tail migrates outwards reaching densities smaller than the minimum density of the disk (at which point we stop tracking it), while the rest is reabsorbed in the disc.

The remaining component, corresponding to the bulk of the disc, has nearly constant entropy of about $\simeq 4.5 \entr$ (the entropy does reach values as high as $\simeq 12 \entr$, but only for fluid elements characterized by a small mass fraction of $\simeq 10^{-5}$ or lower). The constancy of the entropy in the more asymmetric models, as opposed to the trend described above for the near-symmetric ones, can be explained by noting that tidally ejected matter remains cold, and in time undergoes an isothermal expansion which increases its entropy of $\lesssim 3 \entr$. Since the contribution of the tidal component becomes more relevant for increasing mass ratio, the density dependence of the entropy also becomes less noticeable.

Taking into account these observations, for such high-mass ratio models we do not provide a functional form for the $s=s(\rho)$ relation, since it can reasonably be modeled by a constant value.\\

\subsubsection*{Electron fraction.}
\label{sec:electron_fraction}

Matter inside the two \acp{NS} is in neutrinoless, weak equilibrium. However, during the subsequent merger and post-merger phases, the electron fraction in the disk changes due to charged current reactions, both in equilibrium and out-of-equilibrium conditions. The most relevant reactions that we will consider in the following to analyze the electron fraction profiles are the electron capture on protons, the positron capture on neutrons and their inverse reactions:
\begin{align*}
	e^- + p \rightarrow n + \nu &\qquad \rm electron~capture, \\
	e^+ + n \rightarrow p + \bar \nu &\qquad \rm positron~capture, \\
	\nu + n \rightarrow p + e^- &\qquad \rm neutrino~absorption, \\
	\bar \nu + p \rightarrow n + e^+ &\qquad \rm antineutrino~absorption.
\end{align*}

\paragraph{Small mass ratio ($q \lessapprox 1.3$).}

\begin{figure}
	\includegraphics[width=\linewidth]{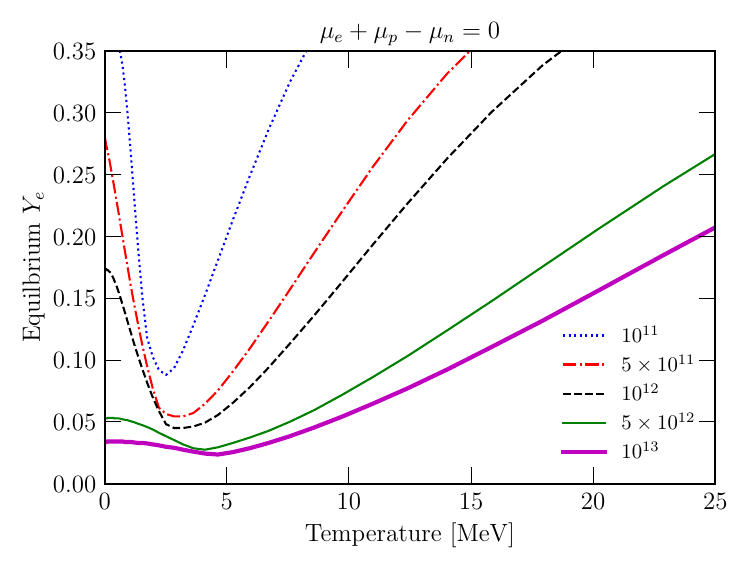}
	\caption{Equilibrium electron fraction for weak reaction in marginally optically thick conditions, i.e. for negligible neutrino fractions, as a function of the temperature and for different rest mass densities ranging between $10^{11}\dens$ and $10^{13}\dens$. The equilibrium is found by solving $\mu_e + \mu_p - \mu_n = 0$ for the BLh \ac{EOS}.}
	\label{fig:equilibrium ye}
\end{figure}

\begin{figure}
	\includegraphics{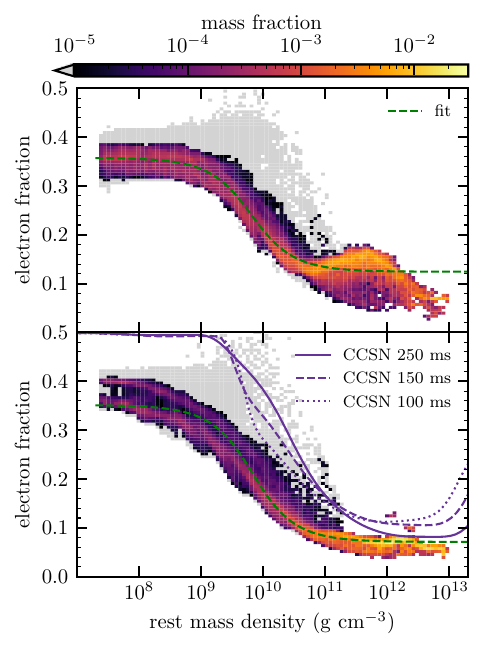}
	\caption{Mass weighted histograms of the rest mass density and electron fraction of the disk for the same representative simulation of \reffig{fig:rmd_entr_chosen} (top) and for the equal mass long-lived \ac{BNS} merger SR simulation with BLh \ac{EOS} (without turbulent viscosity) taken at the end the simulated time (bottom). The fit with \refeq{eq:rmd_ye_fit} is shown using a green dashed line. The purple lines are the density - $\ye$ distributions from the \ac{CCSN} simulation discussed in \ref{sec:discussion} at different post-bounce time.}
	\label{fig:rmd_ye_chosen}
\end{figure}

The mass weighted histogram of \reffig{fig:rmd_ye_chosen} shows the disc's electron fraction distribution with respect to the rest mass density. Most of the mass is characterized by a low electron fraction, with values in the interval $0.1 - 0.2$. At very high densities, $\rho \sim 10^{11} - 10^{13} \dens$, the matter reaches an even lower electron fraction ($\ye \lesssim 0.1$). This is a feature that characterizes all the simulations, regardless of the \ac{EOS} or mass ratio. In particular, the value of the electron fraction can drop below its initial minimum value in the cold, neutrinoless beta-equilibrium \acp{NS}. At such high densities the initial post-merger temperature is $\approx 5-15~\rm MeV$ and decreases to $3-10~\rm MeV$ after the first $\sim 30~\rm ms$ due to the efficient neutrino cooling. These regions are, however, inside the neutrino decoupling regions for both electron neutrinos and antineutrinos. The composition is then set by the equilibrium between neutrino emission and absorption processes. In the limit where the presence of trapped neutrinos is negligible the equilibrium is set by the condition $\mu_{\textnormal{p}} - \mu_{\textnormal{n}} + \mu_{\textnormal{e}} \approx 0$, where $\mu_{\textnormal{n}}$, $\mu_{\textnormal{p}}$, $\mu_{\textnormal{e}}$ are the chemical potentials of neutrons, protons and electrons, respectively.
In Fig.~\ref{fig:equilibrium ye} we present the equilibrium $Y_e$ for the BLh \ac{EOS}. For matter in the rest mass interval $\rho \sim 10^{12} - 10^{13} \dens$ and temperature interval $T \sim 5 - 10~{\rm MeV}$, the equilibrium $\ye$ is always $< 0.1$ and it decreases if $T$ decreases or if $\rho$ increases. This result is not specific for one \ac{EOS}, since it relies on generic features of the matter properties in the relevant temperature and density conditions. Indeed, modeling
the nucleons as a Maxwell-Boltzmann ideal gas of free protons and neutrons, and the electrons as an ultrarelativistic, strongly degenerate ideal gas (under these conditions positrons are suppressed by degeneracy and $Y_e$ becomes a good proxy of the abundance of electrons), the equilibrium conditions can be approximately expressed by:
\begin{equation}
  k_{\rm B}T \ln{\left( \frac{1-\ye}{\ye} \right)} - E_{\rm F}
  \left( 1 - \frac{\pi^2}{6} \left( \frac{k_{\rm B}T}{E_{\rm F}} \right)^2 \right) = 0\, ,
  \label{eq: approximated Ye equilibrium}
\end{equation} 
where for electrons we used $E_{\rm F} = p_{\rm F}c$ and $p_{\rm F} = \hbar (3 \pi^2 \ye \rho/m_b)^{1/3}$. Furthermore we made use of the Sommerfeld lemma to compute the first order correction in $T$ for the electron chemical potential.  For $5~{\rm MeV}\lesssim T \lesssim 10~{\rm MeV}$ and  $ 10^{11} \dens \lesssim \rho \lesssim 10^{13} \dens$ the results obtained by solving Eq.~\eqref{eq: approximated Ye equilibrium} are consistent with the ones presented in Fig.~\ref{fig:equilibrium ye}.

For early enough time, in the $\rho \sim 10^{11} - 10^{12} \dens$ density region, the electron fraction as a function of density shows a local peak. The increase in $\ye$ immediately below $10^{12} \dens$ is mostly due to positron captures happening in hot matter locally shocked or expanding from the innermost part of the disk in a region where electron antineutrinos start to decouple from matter \citep{Endrizzi:2019trv}. Below this density, electron antineutrinos are out of
equilibrium and their capture on protons becomes more effective than positron capture, eventually decreasing $\ye$\footnote{Note that the assessment of the robustness of this feature would require a more realistic neutrino transport
treatment, since the interplay between different neutrino species in the semi-transparent regime (when some species are coupled to matter and others are not) is delicate and it is not obvious that the combination of a leakage and an M0 scheme we employ is able to correctly model all the relevant processes.}.
Nevertheless, this feature is a transient which disappear on a timescale of $~100~\rm ms$, producing an almost monotonic increase between the high and the low density regimes (see bottom panel of \reffig{fig:rmd_ye_chosen}). For large enough time, the conditions inside this part of the disk resemble the innermost conditions inside accretion disk around black holes. Since eventually matter becomes optically thin to neutrinos, the electron fraction reaches an equilibrium condition which is set by the balance between electron and positron captures \citep{Beloborodov:2002af, Foucart:2014nda}. We speculate that on longer timescale the disk will set to a self-regulating stage, in which the neutrino cooling is balanced by the local heat production, for example due to viscous processes \citep{Beloborodov:2002af,Just:2021cls,Siegel:2017jug}.

Going from $10^{11}$ to $10^{9} \dens$, the temperature approximately decreases from $4$ to $2~\rm MeV$. The drop in temperature and density is responsible for a progressive decrease of the relevance of electron and positron captures, while
the decoupling of electron neutrinos from matter favors their absorption on neutron rich matter in free streaming and out-of-equilibrium conditions. The overall effect is a progressive increase of the electron fraction. Below $\rho \sim 10^9 \dens$ the temperature further drops below $\lesssim 2~\rm MeV$. As a consequence, electron and positron captures become negligible. Over time, the electron fraction in the outer part of the disk approaches an equilibrium state around 0.4, determined by the electron (anti)neutrino luminosities and mean energies \citep{Qian:1996xt,Martin:2015hxa}.

In contrast to both long and short-lived \ac{BNS} mergers, the disks in the prompt-collapse category are not irradiated by the neutrinos emitted by the central \ac{NS}. Under these conditions the electron fraction is exclusively determined by the equilibrium between electron and positron captures. We observe that the electron fraction of the disks from prompt-collapse mergers with $q \lesssim 1.3$ increases from 0.04 to $\sim 0.2$ as the density (temperature) decrease from $10^{11}\dens$ ($4~\rm MeV$) to $10^7\dens$ ($1~\rm MeV$). We emphasize however that these values do not represent the equilibrium values of
$\ye$. Indeed, the typical timescales of the electron and positron captures in this thermodynamic conditions range from $\sim 10~\rm ms$ to $10~\rm s$. Our simulations do not extend to such timescales (being shorter than $ 20~\rm ms$
post-merger), therefore we cannot ascertain the ultimate equilibrium value of the electron fraction. This accounts for the consistently lower $\ye$ values observed in the considered disks, which are below the anticipated equilibrium value for neutrino-transparent matter at the same densities and temperatures \cite{Foucart:2014nda}.

Similarly to what is done in \refsec{sec:entropy}, for \ac{BNS} mergers that do not undergo prompt-collapse, it is possible to fit the electron fraction as a function of the rest mass density using a sigmoidal function, \eg an arcotangent:
\begin{equation}\label{eq:rmd_ye_fit}
	\ye(\rho) = Y_{\textnormal{e},0} - \bar{\ye} \arctan(\rho/\rho_0)~.
\end{equation}
The result of the fit is shown in \reffig{fig:rmd_ye_chosen} for the same long-lived simulation of \reffig{fig:rmd_entr_chosen}. The fitting function does not take into account the presence of the local maximum around $\rho \sim 10^{13} - 10^{11} \dens$. This feature is indeed a transient as the neutrino and anti-neutrino diffusion spheres tend to coincide at longer simulation time. The local maximum is indeed disappearing in simulations lasting longer than $100$ ms. However, the presence of this transient in most of the simulations on which we performed the fit shifts the arcotangent plateau at high density to higher electron fraction, decreasing $\bar{\ye}$. The values of the fitted parameters averaged over each category are summarized in \reftab{tab:rmd_ye_fit_par}. The parameters $Y_{\textnormal{e},0}$ and $\bar{\ye}$ are comparable between discs of \ac{BNS} mergers with long-lived and short-lived remnants, indicating that neutrinos are efficient in reprocessing the matter even when the central object collapses in tens of milliseconds. Note that these values too have a qualitative nature, as commented above for the analogue case of the entropy distribution.
\begin{table}
	\caption{Same as in \reftab{tab:rmd_entr_fit_par}, but for the parameters obtained from the least square fits of the $\ye=\ye(\rho)$ relation \refeq{eq:rmd_ye_fit}.}
	\label{tab:rmd_ye_fit_par}
	\begin{tabular}{cc|ccc}
		\toprule
		category &number &$Y_{e,0}$ &$\bar \ye$ &$\rho_0$\\
		&of sim & & &($\times 10^{10} \rm g~cm^{-3}$)\\
		\hline
		long-lived &11 &$0.35 \pm 0.02$ &$0.15 \pm 0.02$ &$0.63 \pm 0.05$\\
		short-lived &8 &$0.33 \pm 0.03$ &$0.14 \pm 0.03$ &$1.2 \pm 0.3$ \\
		\bottomrule
	\end{tabular}
\end{table}
\newline
\paragraph*{High mass ratio ($q \gtrapprox 1.3$).}
Similarly to the entropy for very asymmetric binaries, the electron fraction differs among the tidal and shocked component.

The $\ye$ of the shocked component is determined by the decrease of the electron and positron captures with the temperature and the density, and by the flux of neutrinos, as discussed previously for the $\ac{BNS}$ with $q \lesssim 1.3$. For long and short-lived \ac{BNS} mergers, the electron fraction of the shocked component goes from 0.05 up to 0.4. Only a very small fraction ($< 10^{-5} \mdisc$) of this component can reach values as high as 0.5. If, on the other hand, the remnant undergoes immediate collapse, the maximum $\ye$ is reduced by the lack of neutrino irradiation from the central \ac{NS}.

In all the models, the tidal component is characterized by very neutron-rich matter with very low temperature and $\rho \lesssim 10^{11} \dens$. In this thermodynamic conditions, only electron captures can contribute to the change of the matter composition, reducing the $\ye$ to $0.02 - 0.08$. Further electron conversions are then prevented by the high neutron chemical potential.

Since for high mass ratio the tidal component accounts for most of the disc, disks from high mass ratio \ac{BNS} can be approximately regarded as having constant low $\ye \sim 0.05$.

\section{Discussion}
\label{sec:discussion}

\subsection{Specific angular momentum}
\label{subsec:sam_discussion}
In \citet{Galeazzi:2011nn} the authors proposed a parameterized rotation profile able to describe different classes of differentially rotating \acp{NS}.
Even though rotating \acp{NS} are the intended use case of this model, we apply it here to accretion discs. We have found that this leads to some interesting insights regarding the discs' rotational profile. The profile presented in \citet{Galeazzi:2011nn} is written as:
\begin{equation*}
	g(\Omega) = \frac{\frac{R_0^2}{\Omega_c^{\alpha}} \Omega (\Omega^{\alpha}
    -\Omega_c^{\alpha})}{1- \frac{R_0^2}{\Omega_c^{\alpha}} \Omega^2
    (\Omega^{\alpha} - \Omega_c^{\alpha})} \,,
\end{equation*}
where the quantity on the left-hand side is defined as $j/(1-j\Omega)$. Here $\Omega$ is the angular frequency measured by an
asymptotic inertial observer, while $\alpha$, $R_0$ and $\Omega_c$ are free parameters of the model. In particular, in the Newtonian limit, $\Omega_c$ is the angular frequency around the axis of rotation. The corresponding specific angular momentum of the model $j_G$ is
\begin{equation}\label{eq:galeazzi_jw}
	j_G(\Omega) = \frac{R_0^2}{\Omega_c^{\alpha}} \Omega (\Omega^{\alpha} -
  \Omega_c^{\alpha}) \,.
\end{equation}
Interestingly $\alpha=-1$ and $-4/3$ represent, respectively, the specific angular momentum of a j-const law \citep{1985A&A...146..260E} and of the Kepler law.
\reftab{tab:rmd_amd_galeazzi_fit} shows the one sigma weighted averages of the parameters obtained from a mass-weighted non-linear least square fit on the specific angular momentum of the disk as a function of the angular velocity, taken at the end of each simulation in the sample. It is worth noting that $\alpha \sim -1$ as expected from the results in \refsec{sec:dynamical_properties}. We also fit the Newtonian limit of the specific angular momentum $j_{\rm G}(R)=\Omega_{\rm{G}} R$ as a function of the radius, where $\Omega_G$ is the Newtonian limit of the angular velocity given by the model (equation (11) in \cite{Galeazzi:2011nn}) finding similar results.

\begin{figure}
	\includegraphics{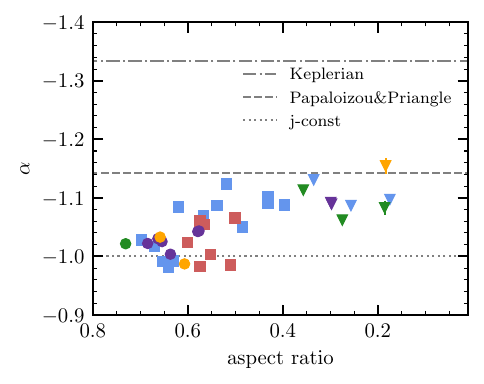}
	\caption{$\alpha$ vs the aspect ratio of the discs. Dotted, dashed and dash-dotted horizontal lines represent the j-const, \citet{Papaloizou:1984} and Keplerian values of $\alpha$.}
	\label{fig:galeazzi_ar_alpha}
\end{figure}

Previous works on \ac{BNS} merger simulations \citep[\eg][]{Ciolfi:2019fie, Hanauske:2016gia} suggested that the Newtonian limit of the angular velocity outside the remnant approach the Kepler law. \citet{Camilletti:2022jms} tried to explain the
relation between $\jdisc$ and $\mdisc$ using the Kepler law to approximate the radial distribution of the angular momentum integrated along $\phi$ and $z$.
The results presented here suggest that the specific angular momentum is instead constant. Since the absolute difference of the angular velocity between the j-const and Kepler laws decreases with the distance from the rotational axis, we believe that the trend of the specific angular momentum is a better discriminant between the two laws. Indeed, in this case the absolute difference between the two models increases with the radius as $r^{2/3}$ and a least-squares fit can easily differentiate between the two cases.
In \reffig{fig:galeazzi_ar_alpha} we plot $\alpha$ vs. the aspect ratio of the discs. We find that the $\alpha$ parameter increases with the aspect ratio, indicating that thinner disks are closer to being Keplerian than thicker ones. In particular, disks characterized by a lower aspect ratio in our sample ($0.4 - 0.2$) have radial distribution of the specific angular momentum broadly compatible with what found by \citet{Papaloizou:1984,Zurek:1986,Nealon:2017}. These works study the redistribution of the angular momentum due to the Papaloizou-Pringle instability, and they find a decrease of the aspect ratio over time and a change in the exponent of the specific angular momentum radial distribution, which tend to a power law whose exponent is $\approx 0.25$, \ie $\alpha \approx 1.14$. This may suggest that \ac{BNS} accretion disks evolve in time from a j-const rotational state to a Keplerian one. However this evolution is likely to take place over long timescales that we cannot investigate due to the limitations of our data sample.

To conclude, note that disks should satisfy the Rayleigh criterion for stability, which states that the specific angular momentum must not decrease outward, \ie $\alpha \lesssim -1$ \citep{Papaloizou:1984}. This condition is fulfilled by most of the disks we study, and in particular $j$-const disks are marginally stable under this criterion.

\subsection{Comparison with disks from \ac{BH}-\ac{NS} mergers}
In \cite{Most:2021ytn}, hereafter \Most, the authors study the properties of the disk formed in \ac{BH}-\ac{NS} mergers. Among their different binary setups, our results are more comparable with the \ac{BH}-\ac{NS} mergers with a non-spinning \ac{BH} (see figure 6 of \Most, $\tilde \chi = 0.00$ case). 
In this scenario, the entropy per baryon has a similar trend compared to what we have found, despite having lower values. This difference is expected since some of the shock's mechanisms described in \refsec{sec:entropy} are possible only in the collision resulting in a \ac{BNS} merger.
The electron fraction of the disk in \Most is usually $\ye \le 0.1$ as in the prompt cases discussed in \refsec{sec:electron_fraction}. Despite this similarity, the simulations in \Most show a local peak in $\ye$ at $\rho \approx 10^9\dens$ that we cannot recognize in our prompt-collapse simulations. Note that the local peak of the electron fraction discussed in \refsec{sec:electron_fraction} for long and short-lived \ac{BNS} mergers is not compatible with what showed in \Most.
Notably, only the \ac{BH}-\ac{NS} mergers with a non spinning \ac{BH} result in a disk exhibiting nearly constant angular momentum within the range $4-7 \times 10^{16}~\rm cm^2~s^{-1}$, consistent with our findings.

\subsection{Accretion rate}
Works that investigated the accretion of the disk onto the central object in the aftermath of a \ac{BNS} merger include \citet{Fernandez:2013tya} (2D long-term simulations); \citet{Siegel:2017jug} and \citet{De:2020jdt} (3D \ac{GRMHD}
simulations); \citet{Fahlman:2022jkh} (pseudo-Newtonian, {MHD} long-term simulations); and \citet{Kiuchi:2022nin} (self-consistent one second long \ac{BNS} merger simulation). They all consider as initial conditions disks characterized by constant specific angular momentum, constant entropy and constant electron fraction around a \ac{BH} of prescribed mass and spin.
The typical accretion rates measured in these works span the range $10^{-3}-1 \msun~\rm s^{-1}$. We find instead higher values of the accretion rate during and after the merger at around $10 \msun~\rm s^{-1}$, also in agreement with the simulation in \citet{Kiuchi:2022nin}. However we observe that on a timescale of $\sim 5~\rm ms$ the accretion rates decreases below $1 \msun~\rm s^{-1}$. The smaller rates measured right from the start in the cited works likely are a consequence of the equilibrium configurations they employ. Indeed, when an initial relaxation phase is included, as in \cite{De:2020jdt}, the measured accretion rate is consistent to the values found in our data

\subsection{Aspect ratio}
In this Section we compare the disk aspect ratio as measured in our analysis (see \refsec{sec:geo_prop}) to the values inferred from the disks presented in the literature. Note however that this quantity is not provided explicitly in most published material. Instead, we extract its value from published 2D plots of discs. To this end we consider a density isocontour in the $xz$ plane around to the typical $\rho_{\rm min}$, \eg $10^8-10^9\dens$. Given this difficulty, the values we obtain are rough estimates at best. Nonetheless they allow to reach some valuable conclusions.
An exception applies to the work of \citet{Kiuchi:2022nin}, for which we directly compute the aspect ratio from the simulation data. In this instance, the disk's aspect ratio at $117~\rm ms$, derived from the isocontour at $10^8\dens$, is 0.30, while it extends to 0.49 for the isocontour taken at $10^9\dens$.
Regarding the \texttt{S\_def} model of \citet{Fernandez:2013tya} at $1.16~\rm s$, the aspect ratio is $\approx 1/4$, as estimated from their Fig. 5. For a density of $\sim 10^6\dens$ the aspect ratio would instead increase to $\approx 1/3$. Fig.~7 of
\cite{Perego:2014fma} also returns an aspect ratio of $\approx 1/3$, while the $10^9\dens$ isocontour of Fig. 4 in \citet{Siegel:2017jug} results in a value of $\approx 0.4$. Finally from Fig.~1 of \citet{Fahlman:2022jkh} we recover an
aspect ratio of $\approx 0.5$ or 1, for the $10^8\dens$ or $10^9\dens$ isocontours, respectively.
The disks in the referenced works are axisymmetric tori around a \ac{BH}. Therefore it is appropriate to compare them to our data from near equal mass mergers with short-lived or prompt-collapsed central objects. We find the aspect ratio of the disks in this subset to be in the range $0.6-0.8$, \ie significantly larger than the disks employed in the literature (the work by \citet{Fahlman:2022jkh} being the only possible exception). While these works cannot be said to employ ``thin'' disks (typically this means $H\simeq10^{-3}$ or lower), better realism might be achieved by setting up initial conditions
with disks that are almost as thick as they are wide, similar to what we find in our data sample.

\subsection{Comparison with Core-Collapse supernova profiles}
The long term evolution of the specific entropy and electron fraction profiles as a function of the rest mass density inside the disk show that both these quantities reach a relatively tight relation, which is relatively insensitive to the properties of the initial binary and of the nuclear \ac{EOS}. This suggests that the shape of these profiles depends on the properties of matter and on the effects of shocks on it in a way that is largely independent from the details of the way in which these profiles are reached. To further test this conclusion, in Fig.~\ref{fig:rmd_entr_chosen} and \ref{fig:rmd_ye_chosen} we compare the specific entropy and the electron fraction profiles inside a representative \ac{BNS} merger simulation with those obtained from spherically symmetric core-collapse supernova simulations of a zero-age main sequence $15\msun$ progenitor star from \citet{Woosley:1995ip} at different times post-bounce, namely 100ms, 150ms and 250ms. In particular, we consider publicly available results obtained by the \texttt{AGILE-BOLTZTRAN} code \citep{Liebendoerfer:2001gu,Liebendoerfer:2002xn}  and published in \citet{Liebendoerfer:2003es}. This simulation included detailed neutrino transport and employed the Lattimer-Swesty \ac{EOS} \citep{Lattimer:1991nc}. In the \ac{CCSN} simulation, the shock wave is launched at bounce from an enclosed mass of $0.5\msun$, where the rest mass density is $\sim 10^{14}\dens$. Afterward, it moves outward shocking radially infalling, low-entropy matter of the stellar layers forming the iron core and the shells above it. As soon as the matter crosses the shock front, the entropy increases. As time passes, even if the radial expansion of the shock stops as it reaches the so-called shock stalling phase, the shock still moves outward in the enclosed mass coordinate due to the continuous mass accretion, reaching lower densities.
Around 100ms (a time which is comparable to our \ac{BNS} merger simulation) the shock is located at $\sim 2 \times 10^8 \dens$ and within a few km the matter density increases by one order of magnitude while increasing also its specific entropy. The latter further increases between $\sim 2 \times 10^8 \dens$ and $\sim 1 \times 10^9 \dens$ due to the effect of neutrino heating. The resulting entropy profile between a few $10^9 \dens$ and $10^{13} \dens$ follows very closely the one observed in the disc. At later times, and in particular at 250ms, a substantial deviation is observed between a few times $10^8 \dens$ and $10^{11} \dens$. This is due to the prolonged neutrino heating and to the contraction of the shock front. 
Such a discrepancy is expected, since our \ac{BNS} merger simulation was evolved only for 100ms and matter in the disk tend to expand rather than to contract.

In the case of the electron fraction, the profiles have a similar
shape, but the ones of the \ac{CCSN} simulation are systematically higher than
the one of the disc. The reason is that in \acp{CCSN} matter with an initial
$Y_e \lesssim 0.5$ and contained inside the stellar core is accreted by the
shock and deleptonizes toward the cold $\beta$-equilibrated conditions that
characterize a \ac{NS}. This condition is achieved passing through the intermediate proto-neutron star phase in which matter is hot and neutrino trapping occurs for high enough matter densities. In \ac{BNS} mergers, the opposite process occurs: cold
$\beta$-equilibrated \ac{NS} matter with $Y_e \lesssim 0.1$ is heated and
decompressed inside the disc, and it tends to leptonize, at least for low enough densities. At late enough time, in the high density part of the profile 
($\rho \gtrsim 10^{12} \dens$), the \ac{CCSN} profile approaches the one inside the disc, as weak reaction equilibrium is achieved. For lower densities the visible discrepancy is due to the out-of-equilibrium character of the weak reactions, which prevents the profiles from reaching a state that has completely lost memory of its history. Moreover, such an equilibrium depends also on the neutrino irradiation, which has different features in \acp{CCSN} and \ac{BNS} mergers.

\section{Conclusions}
\label{sec:conclusions}

We have studied the geometrical, dynamical and thermodynamical properties of \samplesize disks from numerical relativity simulations of \ac{BNS} mergers, classified by the fate of remnant: \longlived long-lived, \shortlived short-lived and \prompt prompt-collapsed.
Most of our simulations are targeted to the \ac{BNS} merger GW170817, with $M_{\rm chirp} = 1.18\msun$. A subset of 6 simulations are targeted to GW190425, with $M_{\rm chirp} = 1.44\msun$.

We found that \ac{BNS} accretion disks are remarkably thick. In particular, the aspect ratio of the disks from mergers that do not undergo prompt-collapse decreases with the mass ratio, going from $\approx 0.8$ to below 0.3, while disks from prompt-collapsed mergers span the range $0.4-0.15$. Such a large aspect ratio reflects the significant thermal support inside the disk during the first tens of milliseconds after merger.

The mass and angular momentum of the disks span a broad range of values, going from $5 \times 10^{-4}$ to $0.3 \msun$, for the mass and from $2 \times 10^{13}$ to $10^{16} \am$ for the angular momentum. We have found that the specific angular momentum is almost constant in any of the disk in our sample, taking values between 3 and $5~\times 10^{16}\sam$. This is also confirmed by the distribution of the specific angular momentum with the angular velocity, which is compatible with the so called $j$-const law.

In the first $10-15$ ms after merger, disks where a central massive \ac{NS} is present show an outflow of matter at any radial distance from the remnant, which decrease with time and radius from a maximum of $10 \msun~\rm s^{-1}$. A persistent accretion only occurs when the central object collapse to a \ac{BH}, with an initial accretion of $10 \msun~\rm s^{-1}$. After an initial transient phase which lasts $\sim 5~\rm ms$, both ejection and accretion rates decrease to $1 \msun~\rm s^{-1}$, similar to the values found in many works of long-term disk evolution, where the initial disks are considered as equilibrium tori.

The specific entropy in the disk has different behaviors depending on the mass ratio of the binary. For small mass ratios ($\lessapprox 1.3$), most on the matter in the disk spans the rather limited entropy range, with entropy of $4-8\entr$ in the density range $10^{10}-10^{13}\dens$. It must be stressed that, at lower densities, the entropy of a non-negligible fraction of the disk increases to $15-20\entr$. We have found that the entropy is distributed around a sigmoidal function of the rest mass density, which can be satisfactorily modeled using a modified arcotangent. For higher mass ratios, the disks
decompose in a tidal and a shocked components. In this case the bulk of the disk can be regarded as approximately isentropic.

Similar to the specific entropy, also the behavior of the electron fraction inside the disk changes according to the mass ratio. For $q \lesssim 1.3$, in the high density region ($10^{11}-10^{13}\dens$) the matter is neutron rich ($\ye \approx 0.1-0.2$). At lower density the electron fraction increases to $\approx 0.4$ and a negligible amount of matter, with respect to the total disc mass, reaches even higher $\ye$. The distribution of the electron fraction with the density follows a sigmoidal function, for which we provide a fit in terms of an arcotangent. At higher mass ratios the electron fraction, as the entropy, is nearly constant and very low.

We note that the simulations that constitute our sample and on which we base our analysis do not include some physical input that might affect our findings. In particular, they do not take into account the presence magnetic fields, and the treatment of neutrino interactions and transport is somewhat simplistic, due to the use of the \texttt{Leakage + M0} (see \refsec{sec:code}). Having access to more realistic datasets that model these processes might have an impact on \eg the spatial distribution of matter in the disks (affecting the aspect ratio and rotational profile) and/or their composition and thermodynamics.

In this work however, our objective is to provide a comprehensive qualitative description of \ac{BNS} accretion discs, until now missing in the literature on the subject. The effects mentioned above are going to have only rather limited quantitative effects, such as varying our estimates for disks parameters to the level of a few percents. Therefore we are confident that the description we provide is qualitatively realistic and sound, and very unlikely to dramatically change by more realistic simulation setups.

In light of these considerations, we think that the characterization of \ac{BNS} accretion disks that we provide can be useful to the scientific community. First of all, the structure of such disks is interesting in its own right and it has
not yet systematically been studied in the literature. Furthermore, as mentioned in previous Sections, many works that perform simulations of \ac{BNS} accretion disks employ initial conditions that lack in realism, particularly regarding the
setup of the thermodynamic state and composition of the discs. Such simulations could achieve greater realism by employing disks models that are closer to the specifics we have provided.

Finally, we have uncovered some properties of the accretion disks that deserve further investigation in their own right, \eg the mechanism by which their rotational profile achieves a configuration of constant specific angular momentum and whether they evolve towards a Keplerian profile on secular timescales. We however leave this investigation for future work.

\section*{Data availability}
Data generated for this study will be made available upon reasonable request to the corresponding authors.

\begin{acknowledgments}
  The Authors thank Dennis Verra for preliminary work on the project.
  The Authors acknowledge the INFN and Virgo for the usage of computing and storage resources through the tullio server in Torino. 
  The work of AP is partially funded by the European Union under NextGenerationEU. PRIN 2022 Prot. n. 2022KX2Z3B.
  DR acknowledges funding from the U.S. Department of Energy, Office of Science, Division of Nuclear Physics under Award Number(s) DE-SC0021177, DE-SC0024388, and from the National Science Foundation under Grants No. PHY-2011725,   PHY-2020275, PHY-2116686, and AST-2108467.
  SB acknowledges funding from the EU Horizon under ERC Consolidator Grant, no. InspiReM-101043372.
  FMG acknowledges funding from the Fondazione CARITRO, program Bando post-doc 2021, project number 11745.
  
  Simulations were performed on the machines Bridges2, Expanse (NSF XSEDE allocation TG-PHY160025), Frontera (NSF LRAC allocation PHY23001), and Perlmutter. 
  AP acknowledges PRACE for awarding him access to Joliot-Curie at GENCI@CEA (project: 2019215202, allocation RA5202).  
  This research used resources of the National Energy Research Scientific Computing Center, a DOE Office of Science User Facility supported by the Office of Science of the U.S.~Department of Energy under Contract No.~DE-AC02-05CH11231.
  
  Numerical calculations have also been made possible through a CINECA-INFN agreement, providing access to resources on the MARCONI/Galileo100/LEONARDO machines at CINECA. 
\end{acknowledgments}

\bibliographystyle{apsrev4-1}

\appendix\label{sec:appendix}

\section{Correlations in the density profile fitting parameters}
\label{app:rmd_rz}

\begin{figure}
	\centering
	\includegraphics{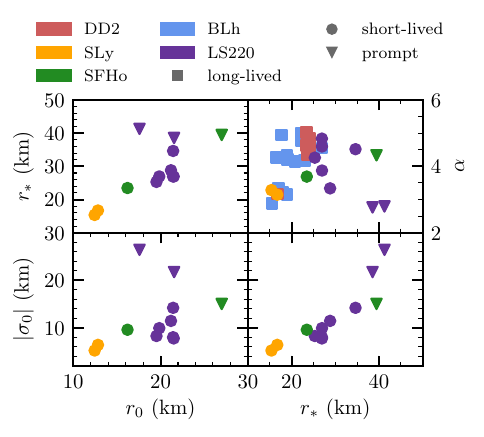}
	\caption{Parameters obtained from the fit described in \refpar{eq:rmd0_rz}.}
	\label{fig:rmd_rz0}
\end{figure}

\begin{figure}
	\centering
	\begin{subfigure}[b]{\columnwidth}
		\includegraphics[width=\columnwidth]{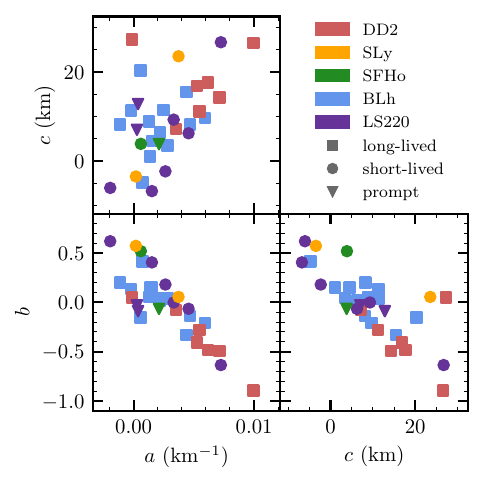}
	\end{subfigure}
	\begin{subfigure}[b]{\columnwidth}
		\includegraphics[width=\columnwidth]{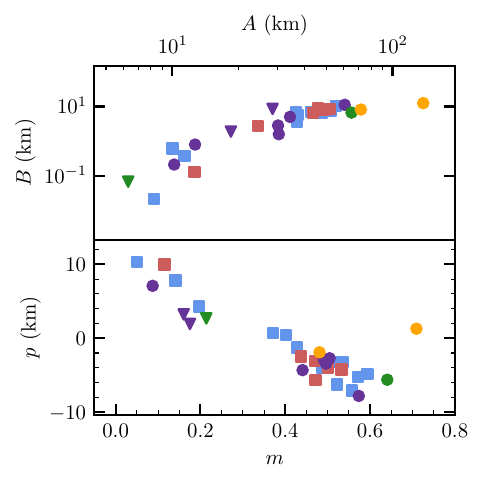}
	\end{subfigure}
	\caption{Parameters obtained from the fit described in \refpar{par:rmd_rz}
    (Eqs.~\ref{eq:rmd_rz_fit_param}).}
	\label{fig:rmd_rz_corr}
\end{figure}

Here we report and briefly discuss the values of the parameters obtained from the fitting procedure described in \refsec{par:rmd_rz}, for each simulation to which it was applied. \reftab{tab:rmd_rz_parameters} reports the parameters of
\refeq{eq:rmd0_rz} and Eqs.~(\ref{eq:rmd_rz_fit_param}). Note that $\alpha$ is computed as $\alpha = 2 r_* (r_* - r_0) / \sigma_0^2$ for short-lived and prompt-collapse \ac{BNS} mergers (values in parenthesis), while it is a free parameter of the fit for long-lived \ac{BNS} mergers.

\begin{sidewaystable*}
	\centering
	\caption{Parameters obtained from the fitting procedure described in \refsec{par:rmd_rz}. From left to the right: category according to the fate of remnant (see \refsec{sec:post-processing}), \ac{EOS}, total initial gravitational mass of the isolated neutron stars $M_{\rm tot}$, mass ratio $q$, inclusion of turbulent viscosity, resolution of the finest grid, disk mass $\mdisc^{\rm end}$, angular momentum $\jdisc^{\rm end}$, values of the parameters $\sigma_0$, $r_0$, $r_*$, $\alpha$ of \refeq{eq:rmd0_rz} and $m$, $p$, $a$, $b$, $c$, $A$, $B$ of \refeq{eq:rmd_rz_fit}.}
	\label{tab:rmd_rz_parameters}
	\begin{tabular}{cccccc|cc|cccc|cc|ccc|cc}
		\toprule
		Category &EOS &$M_{\rm tot}$ &$q$ &GRLES &res
		&$\mdisc^{\rm end}$ &$\jdisc^{\rm end}$
		&$\sigma_0$ &$r_0$ &$r_*$ &$\alpha$ &$m$ &$p$ &$a$ &$b$ &$c$ &$A$ &$B$ \\
		& & & & & &$\msun$ &$\msun^2$n	&km &km &km & & &km &$\rm 10^{-3}~km^{-1}$ & &km &km &km \\
		\hline	long lived &BLh &2.728 &1 &\cmark &HR &0.2081 &1.5398 &$-$ &$-$ &22.30 &4.77 &0.54 &-3.19 &4.70 &-0.14 &8.17 &52.45 &7.72 \\
		long lived &BLh &2.728 &1 &\cmark &SR &0.1328 &1.0831 &$-$ &$-$ &22.16 &5.00 &0.56 &-7.10 &2.80 &0.04 &3.45 &55.35 &10.31 \\
		long lived &BLh &2.728 &1 &\xmark &HR &0.1139 &0.8875 &$-$ &$-$ &16.40 &4.27 &0.57 &-5.28 &2.48 &0.04 &11.43 &36.76 &3.71 \\
		long lived &BLh &2.730 &1 &\xmark &SR &0.0955 &0.7402 &$-$ &$-$ &17.67 &4.94 &0.52 &-6.25 &0.73 &0.42 &-4.78 &36.49 &6.87 \\
		long lived &BLh &2.730 &1 &\xmark &LR &0.1253 &0.9321 &$-$ &$-$ &18.99 &4.33 &0.59 &-4.81 &5.91 &-0.21 &9.64 &50.30 &7.43 \\
		long lived &BLh &2.765 &1.34 &\xmark &HR &0.2024 &1.5797 &$-$ &$-$ &17.00 &3.35 &0.40 &0.42 &1.37 &0.15 &1.02 &37.32 &5.64 \\
		long lived &BLh &2.765 &1.34 &\xmark &SR &0.1664 &1.3716 &$-$ &$-$ &19.09 &4.20 &0.49 &-4.06 &1.52 &0.15 &4.53 &42.54 &7.06 \\
		long lived &BLh &2.765 &1.34 &\xmark &LR &0.2268 &1.8612 &$-$ &$-$ &20.80 &4.14 &0.51 &-3.25 &2.20 &0.02 &6.43 &47.62 &6.77 \\
		long lived &BLh &2.803 &1.54 &\xmark &HR &0.2454 &2.0294 &$-$ &$-$ &18.97 &3.15 &0.05 &10.34 &-1.11 &0.20 &8.21 &10.02 &0.62 \\
		long lived &BLh &2.803 &1.54 &\xmark &LR &0.2594 &2.2244 &$-$ &$-$ &22.95 &4.17 &0.43 &-1.28 &4.41 &-0.33 &15.51 &46.75 &7.23 \\
		long lived &BLh &2.837 &1.66 &\cmark &HR &0.2538 &2.1919 &$-$ &$-$ &15.52 &2.89 &0.20 &4.28 &0.59 &-0.15 &20.32 &11.38 &0.38 \\
		long lived &BLh &2.837 &1.66 &\cmark &SR &0.2650 &2.2929 &$-$ &$-$ &18.04 &3.21 &0.14 &7.84 &-0.20 &0.14 &11.32 &8.27 &0.02 \\
		long lived &BLh &2.837 &1.66 &\cmark &LR &0.2439 &2.2047 &$-$ &$-$ &26.97 &4.55 &0.37 &0.68 &1.29 &0.05 &8.91 &36.86 &4.19 \\
		long lived &DD2 &2.728 &1 &\xmark &SR &0.1811 &1.4914 &$-$ &$-$ &23.39 &5.01 &0.47 &-5.64 &3.51 &-0.08 &7.16 &45.90 &8.82 \\
		long lived &DD2 &2.728 &1 &\xmark &LR &0.2117 &1.7140 &$-$ &$-$ &23.33 &4.63 &0.50 &-4.03 &6.17 &-0.48 &17.69 &51.63 &8.60 \\
		long lived &DD2 &2.732 &1.10 &\xmark &LR &0.2349 &1.9157 &$-$ &$-$ &24.13 &4.85 &0.44 &-2.48 &9.93 &-0.90 &26.59 &24.55 &2.78 \\
		long lived &DD2 &2.733 &1.11 &\xmark &LR &0.2582 &2.0782 &$-$ &$-$ &24.18 &4.56 &0.50 &-3.11 &7.11 &-0.49 &14.28 &43.49 &6.56 \\
		long lived &DD2 &2.740 &1.19 &\xmark &LR &0.2557 &2.0764 &$-$ &$-$ &23.48 &4.36 &0.53 &-4.27 &5.25 &-0.41 &16.82 &48.61 &8.08 \\
		long lived &DD2 &2.742 &1.20 &\xmark &LR &0.2530 &2.0934 &$-$ &$-$ &23.89 &4.51 &0.47 &-3.11 &5.46 &-0.28 &11.20 &46.21 &7.91 \\
		long lived &DD2 &2.880 &1.67 &\cmark &SR &0.2753 &2.5445 &$-$ &$-$ &16.72 &3.17 &0.12 &10.00 &-0.12 &0.05 &27.29 &12.63 &0.13 \\
		\hline
		short lived &LS220 &2.728 &1 &\cmark &SR &0.0502 &0.3901 &8.29 &19.53 &25.32 &(4.26) &0.50 &-3.43 &4.56 &-0.06 &6.25 &34.24 &5.04 \\
		short lived &LS220 &2.728 &1 &\cmark &LR &0.1605 &1.2549 &14.23 &21.44 &34.63 &(4.52) &0.50 &-2.71 &7.25 &-0.64 &26.71 &30.47 &1.60 \\
		short lived &LS220 &2.728 &1 &\xmark &SR &0.0631 &0.4951 &7.78 &21.50 &26.94 &(4.83) &0.44 &-4.32 &2.64 &0.18 &-2.30 &12.69 &0.81 \\
		short lived &LS220 &2.728 &1 &\xmark &LR &0.0697 &0.5450 &8.04 &21.44 &26.95 &(4.60) &0.57 &-7.81 &3.33 &-0.00 &9.32 &60.65 &11.17 \\
		short lived &LS220 &2.737 &1.16 &\cmark &SR &0.1238 &0.9947 &9.95 &19.86 &26.96 &(3.87) &0.49 &-2.88 &1.52 &0.41 &-6.75 &30.22 &2.85 \\
		short lived &LS220 &2.781 &1.43 &\cmark &LR &0.1855 &1.6314 &11.47 &21.19 &28.82 &(3.34) &0.09 &7.10 &-1.94 &0.62 &-6.02 &10.20 &0.22 \\
		short lived &SFHo &2.735 &1.13 &\xmark &SR &0.0775 &0.5902 &9.61 &16.22 &23.48 &(3.69) &0.64 &-5.62 &0.60 &0.52 &3.88 &65.16 &6.68 \\
		short lived &SLy4 &2.728 &1 &\xmark &SR &0.0417 &0.3136 &5.24 &12.46 &15.39 &(3.29) &0.71 &1.29 &3.73 &0.06 &23.56 &137.87 &12.48 \\
		short lived &SLy4 &2.735 &1.13 &\xmark &SR &0.0687 &0.5282 &6.43 &12.84 &16.73 &(3.15) &0.48 &-1.90 &0.19 &0.57 &-3.45 &71.85 &8.16 \\
		\hline
		prompt &LS220 &2.837 &1.66 &\cmark &LR &0.1185 &1.2242 &26.44 &17.59 &41.26 &(2.79) &0.18 &1.93 &0.37 &-0.09 &12.82 &28.54 &8.48 \\
		prompt &LS220 &2.837 &1.66 &\xmark &LR &0.1155 &1.1760 &21.76 &21.55 &38.53 &(2.77) &0.16 &3.23 &0.28 &-0.03 &7.04 &18.45 &1.91 \\
		prompt &SFHo &2.837 &1.66 &\cmark &SR &0.0906 &0.8925 &15.04 &27.00 &39.44 &(4.33) &0.21 &2.68 &2.11 &-0.07 &3.80 &6.31 &0.07 \\
		\bottomrule
	\end{tabular}
	\vspace{-.3\textheight}
\end{sidewaystable*}

It is apparent that some of the parameters may not be independent from each other. In \reffig{fig:rmd_rz0} we show the parameters of \refeq{eq:rmd0_rz} as functions of each other. Clearly $r_*$ and $|\sigma_0|$ show some hint of correlation. A similar observation holds for \reffig{fig:rmd_rz_corr}. One can see that, \eg parameter $b$ appears to be a linear function of parameter $a$. The same could be said of $p$ with respect to $m$ and other couples of parameters. This suggests that the fit formulas proposed in \refsec{par:rmd_rz} are to some extent redundant and could be simplified. However we leave the investigation of this possibility to future work.

\end{document}